\documentclass[pre,aps,twocolumn,superscriptaddress,showpacs]{revtex4-1}
\usepackage[latin9]{luainputenc}
\usepackage{color}
\usepackage{units}
\usepackage{mathrsfs}
\usepackage{amsmath}
\usepackage{amssymb}
\usepackage{graphicx}
\usepackage{wasysym}
\usepackage[unicode=true,
 bookmarks=false,
 breaklinks=false,pdfborder={0 0 1},backref=false,colorlinks=true]
 {hyperref}
\hypersetup{
 urlcolor=blue,anchorcolor=blue,citecolor=blue,filecolor=blue,linkcolor=blue,menucolor=blue}

\makeatletter
\usepackage{epsfig}
\usepackage{subfigure}
\usepackage{graphicx}
\usepackage{textcomp}
\usepackage{url}
\usepackage{float}
\usepackage{color}
\usepackage{cases}
\usepackage[normalem]{ulem}

\usepackage{color}

\usepackage{epsfig}
\usepackage{textcomp}
\usepackage{epstopdf}

\makeatother

\begin{document}
\global\long\def\erf{\operatorname{erf}}

\global\long\def\erfc{\operatorname{erfc}}

\global\long\def\sech{\operatorname{sech}}

\title{Large fluctuations of a Kardar-Parisi-Zhang interface on a half-line:
the height statistics at a shifted point}

\author{Tomer Asida}
\email{tomer.asida@mail.huji.ac.il}

\author{Eli Livne}
\email{livne@phys.huji.ac.il}

\author{Baruch Meerson}
\email{meerson@mail.huji.ac.il}

\affiliation{Racah Institute of Physics, Hebrew University of Jerusalem, Jerusalem
91904, Israel}

\begin{abstract}
We consider a stochastic interface $h(x,t)$, described by
the $1+1$ Kardar-Parisi-Zhang (KPZ) equation on the half-line $x\geq0$
with the reflecting boundary at $x=0$. The interface is initially
flat, $h(x,t=0)=0$. We focus on the short-time probability distribution
$\mathcal{P}\left(H,L,t\right)$ of the height $H$ of the
interface at point $x=L$. Using the optimal fluctuation
method, we determine the (Gaussian) body of the distribution and the
strongly asymmetric non-Gaussian tails. We find that the slower-decaying
tail scales as $-\sqrt{t}\,\ln\mathcal{P}\simeq\left|H\right|^{3/2}f_{-}\left(L/\sqrt{\left|H\right|t}\right)$,
and calculate the function $f_{-}$ analytically. Remarkably,
this tail exhibits a first-order dynamical phase transition at a critical
value of $L$, $L_{c}=0.60223\dots\sqrt{\left|H\right|t}$. The transition
results from a competition between two different   fluctuation
paths of the system. The faster decaying tail scales as $-\sqrt{t}\,\ln\mathcal{P}\simeq|H|^{5/2}f_{+}\left(L/\sqrt{|H|t}\right)$.
We evaluate the function $f_{+}$ using a specially developed
numerical method, which involves solving a nonlinear second-order
elliptic equation in Lagrangian coordinates. The faster-decaying tail
also involves a sharp transition, which occurs at a critical value
$L_{c}\simeq2\sqrt{2|H|t}/\pi$. This transition is similar to the
one recently found for the KPZ equation on a ring, and we believe that it
has the same fractional order $5/2$. It is smoothed, however, by small diffusion
effects.
\end{abstract}
\maketitle

\section{Introduction}
\label{Sec: introduction}

The Kardar-Parisi-Zhang (KPZ) equation \citep{KPZ}
is a paradigmatic model of non-equilibrium stochastic growth. It describes
the evolution of the height $h(x,t)$ of a growing surface at the
point $x$ of a substrate at time $t$:
\begin{equation}
\partial_{t}h=\nu\partial_{x}^{2}h+\frac{\lambda}{2}\left(\partial_{x}h\right)^{2}+\sqrt{D}\,\xi(x,t).\label{eq:KPZ_dimensional}
\end{equation}
The Gaussian noise $\xi(x,t)$ has zero mean and is delta-correlated
in space and in time:
\begin{equation}
\langle\xi(x_{1},t_{1})\xi(x_{2},t_{2})\rangle=\delta(x_{1}-x_{2})\delta(t_{1}-t_{2}).\label{eq: correlator}
\end{equation}
Without loss of generality we assume that the nonlinearity coefficient
$\lambda$ is negative \citep{signlambda}. The KPZ dynamics in 1+1
dimension have been studied in detail in numerous works. At long times,
the interface width grows as $t^{1/3}$, and the lateral correlation
length grows as $t^{2/3}$. The exponents $1/3$ and $2/3$ are the
hallmark of a whole universality class of the 1+1 dimensional non-equilibrium
growth \cite{HHZ,Barabasi,Krug,Corwin,QS,S2016,Takeuchi2017}. A sharper
characterization of the KPZ growth is achieved by studying, in a proper
moving frame \cite{displacement}, the full probability distribution
$\mathcal{P}\left(H,L,t\right)$ of the surface height at a specified
point $x=L$ at time $t$. In a translationally invariant system one
can always set $L=0$ and deal with $H=h\left(x=0,t\right)$. Surprisingly,
the form of the distribution, $\mathcal{P}\left(H,t\right)$, at all
times, depends on the initial interface shape $h\left(x,t=0\right)$,
see Refs. \cite{QS,S2016,Takeuchi2017} for recent reviews.

Traditionally (and justifiably), most of the interest in the KPZ equation
has been in the long-time regime, $t\gg\nu^{5}/(D^{2}\lambda^{4})$ and ensuing universality.
More recently the short-time
behavior, $t\ll\nu^{5}/(D^{2}\lambda^{4})$, of the one-point
height distribution $\mathcal{P}\left(H,t\right)$ has started attracting
interest \cite{KK2007,KK2008,KK2009,MKV}. This interest stemmed from a discovery of unexpected scaling
behaviors of the distribution \emph{tails}, which describe atypically
large fluctuations of height. For stationary (random) initial condition, a second-order dynamical phase transition
was discovered \cite{Janas2016}, and a Landau theory of this short-time phase transition has been formulated \cite{SKM2018}.  As of today, \emph{exact} short-time height
distributions have been found for infinite systems with droplet \citep{DMRS},
stationary \citep{LeDoussal2017} and flat \cite{SM2018} initial
conditions. For several other initial conditions, asymptotics of the
distribution \emph{tails} have been calculated. Quite often the tails, found at short times, persist (at sufficiently
large $H$) at arbitrary times \cite{MKV,SMP,MSchmidt2017,Corwinetal2018,Krajenbrinketal2018}.

Another recent development concerns the role of system boundaries.
Ref. \cite{SMS2018} studied the short-time behavior of $\mathcal{P}\left(H,t\right)$
on a ring of length $2L$ and uncovered a whole  phase diagram
of different scaling behaviors of the distribution in the $(L/\sqrt{t},H)$
plane. Other papers have dealt with a more basic setting of a half-line
$x\geq0$, both at long \cite{GueudreLeDoussal2012,Borodin2016,Barraquand2017,CorwinShen2018,ItoTakeuchi2018,Corwinetal2018,Krajenbrink2018}
and at short \cite{GueudreLeDoussal2012,SM2018,Krajenbrink2018,MV2018}
times.

\begin{figure}
\centering{}\includegraphics[scale=0.6]{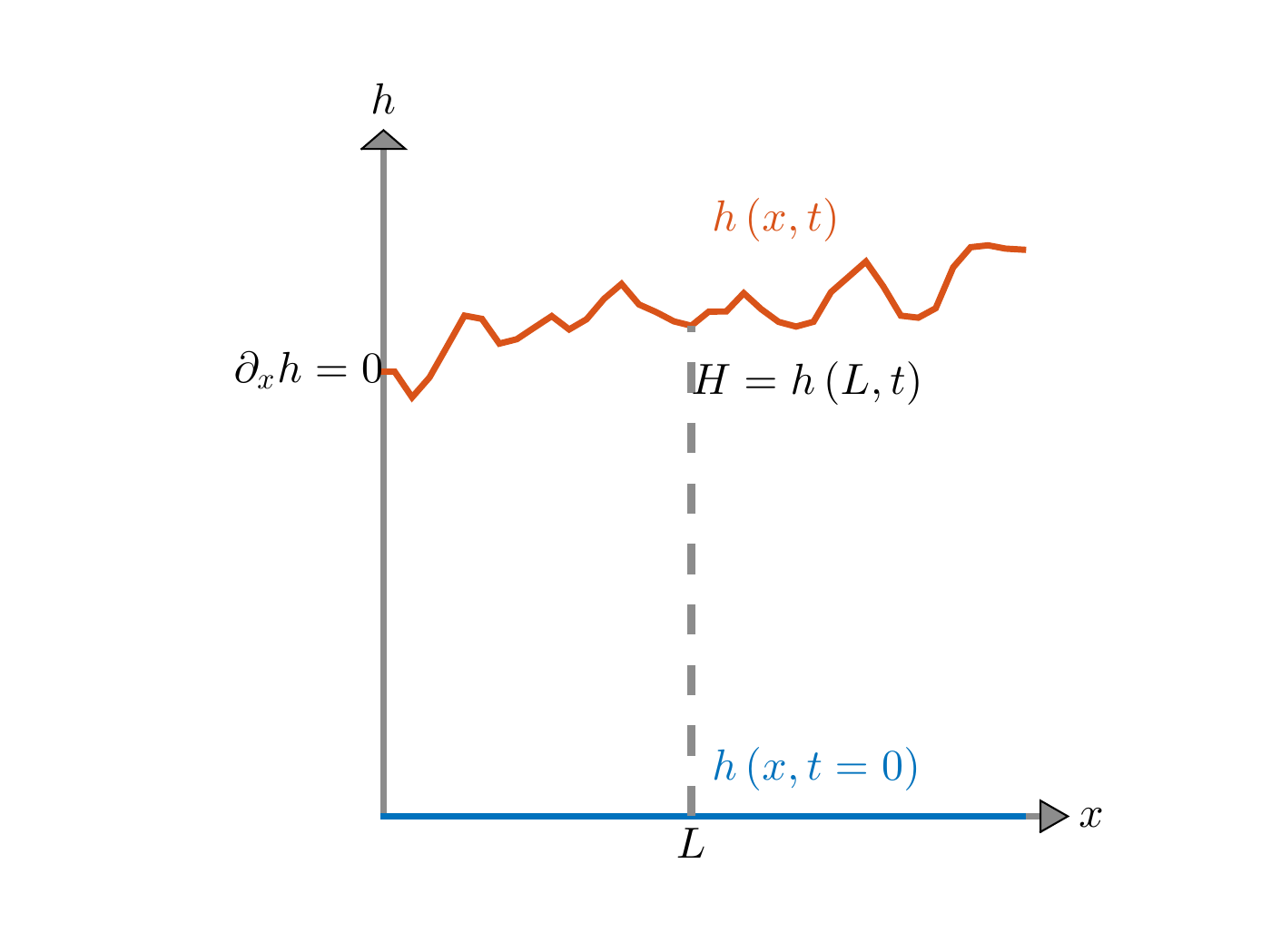}
\caption{A schematic of the problem.}
\label{fig: schematic of problem}
\end{figure}

As in the recent paper \cite{MV2018}, here we will study a KPZ interface
on a half-line $x\geq0$. In Ref. \cite{MV2018} the boundary condition at $x=0$ specified a constant non-zero slope $\partial_{x}h(x=0,t)$ and thus introduced
an additional, deterministic, driving of the initially flat interface. In this work we will assume a reflecting boundary,
$\partial_{x}h(x=0,t)=0$, and an initially flat interface, $h(x,t=0)=0$,
$x\geq0$, but condition the KPZ process on reaching a height $H$
at time $t$ at a \emph{shifted} point of the substrate: $h(x=L,t)=H$. Similarly to the ring problem \cite{SMS2018} (see also Ref. \cite{SKM2018}), the
shifted point introduces a nontrivial additional parameter $L$ into the
problem. In contrast to the ring problem, the additional parameter $L$ keeps the system (half-)infinite. A remote analog of the additional parameter $L$ is the magnetic field in the Ising model of phase transitions. The magnetic field breaks the symmetry between the two phases, whereas the additional length $L$ breaks the mirror symmetry of the optimal interface histories around $x=L$ and leads to new dynamical phase transitions, as we demonstrate below.
A schematic of the problem is shown in Fig. \ref{fig: schematic of problem}.  We will limit ourself to the short-time regime.

The particular case $L=0$ is well understood by applying symmetry arguments to the known solution for the infinite system \cite{SM2018}. For $L=0$
one observes, at short times, a scaling behavior
\begin{equation}
-\ln\mathcal{P}(H,L=0,t)\simeq\frac{\nu^{5/2}}{D\left|\lambda\right|^{2}\sqrt{t}}
\,s_{0}\left(\frac{\left|\lambda\right|H}{\nu}\right)\label{eq: simplescaling}
\end{equation}
with a simple relation
\begin{equation}
s_{0}\left(\frac{\left|\lambda\right|H}{\nu}\right)=\frac{1}{2}\,s_{\text{full}}\left(\frac{\left|\lambda\right|H}{\nu}\right)
\label{eq: relation}
\end{equation}
between the large deviation functions
of the half-line and the full-line problems \cite{SM2018}. For $L>0$, the large deviation function $s(H,L,t)$ is unknown, and it will be in the focus of our
attention.

Our approach to this problem relies on the optimal fluctuation method
(OFM), also known as weak-noise theory, or instanton method.
The OFM has been used in many papers on the KPZ equation and related
systems \citep{Mikhailov1991,GurarieMigdal1996,Fogedby1998, Fogedby1999,Nakao2003, KK2007,KK2008,KK2009,Fogedby2009,MKV,KMSparabola,Janas2016,MSchmidt2017,MSV_3d, SMS2018,SKM2018,MV2018}.
The OFM derives from a path-integral formulation of the conditioned
stochastic process. For an effectively weak noise, one can evaluate
the path integral by the Laplace's method. This procedure leads to
a variational problem. Its least-action solution is the optimal path \textendash{}
the most probable history of the conditioned stochastic process. The
``classical action\char`\"{} along the optimal path, yields $\mathcal{P}$
up to a pre-exponential factor. As we show here, the short-time probability
distribution $\mathcal{P}$ exhibits the following scaling:
\begin{equation}
-\ln\mathcal{P}\left(H,L,t\right)\simeq\frac{\nu^{5/2}}{D\left|\lambda\right|^{2}\sqrt{t}}s\left(\frac{\left|\lambda\right|H}{\nu},\frac{L}{\sqrt{\nu t}}\right).\label{eq: physical variables P(H,L,t)}
\end{equation}
This scaling behavior is the same as in the ring problem \cite{SMS2018},
but the large deviation function $s$ is of course different. The OFM makes it
clear that, as $L\to\infty$, the boundary condition at $x=0$ becomes
irrelevant, and $\mathcal{P}(H,L,t)$ should approach the full-line
distribution. As we will show here, $s$ increases {[}and, therefore,
$\mathcal{P}(H,L,t)$ decreases{]} monotonically with an increase
of $L$, interpolating between one half and the full value of $s_{\text{full}}(\lambda H/\nu)$.
This ``interpolation\char`\"{}, however, looks very differently in
the Gaussian body of the distribution (that is, for relatively small
$|H|$) and in its tails.

In the Gaussian regime, the $L$-dependence of $s$ is smooth at all
$L$. For the $H\to-\infty$ tail (to remind the reader, we assume $\lambda<0$) we find the following scaling behavior
\begin{equation}
-\ln\mathcal{P}(H,L,t)\simeq\frac{|H|^{3/2}}{\sqrt{t}}\,f_{-}\left(\frac{L}{\sqrt{|H|t}}\right),\label{eq: H to -=00005Cinfscaling}
\end{equation}
where, for brevity, we suppressed the constants $\nu$, $D$ and $\lambda$.
We were able to calculate the function $f_{-}$ analytically, see Eq. \eqref{eq: negative tail scaling funcion f_} and Fig. \ref{fig: negative tail action}. Remarkably, it exhibits
a first-order phase transition \textendash{} a discontinuity of its
first derivative \textendash{} at a critical value of $L$, $L_{c}=0.60223\dots\sqrt{\left|H\right|t}$.
At $L>L_{c}$, the large deviation function $s$ is independent of
$L$ and equal to its value for the full line. As we show here, the first-order transition
results from a competition between two different OFM solutions.

\begin{figure}
\centering{}\includegraphics[scale=0.45]{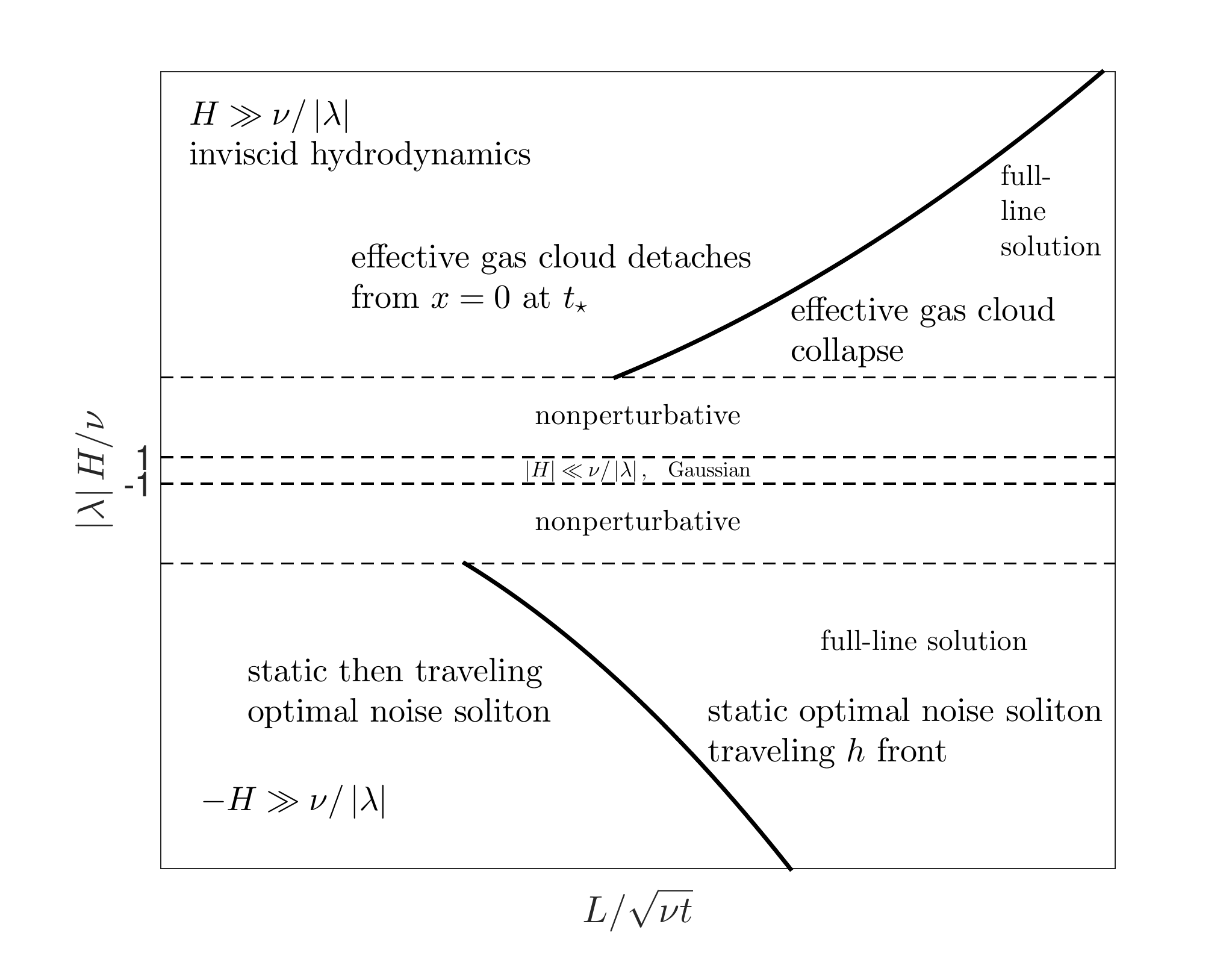}\caption{A phase diagram of the system in the $\left(L/\sqrt{\nu t},\,\left|\lambda\right|H/\nu\right)$
plane. The solid lines and dashed lines denote sharp and smooth transitions,
respectively. For sufficiently large $L$ the half-line system behaves
as the full-line system. Typical (small) height fluctuations are Gaussian,
see Sec. \ref{Sec: lineartheory}. For large negative $H$ the solution
involves a static or traveling optimal noise soliton and is described
in Sec. \ref{Sec: negativetail}. For large positive $H$
the solution is approximately inviscid, and describes a hydrodynamic collapse
of an effective gas cloud, see Sec. \ref{Sec: positivetail}. Both sharp
transition lines are given by $L/\sqrt{t}\sim\sqrt{\left|H\right|}$. ``Nonperturbative" denotes intermediate regions where there is no analytical theory.}
\label{fig: phase diagram}
\end{figure}

For the $ H\to \infty$ tail the scaling behavior of $\mathcal{P}(H,L,t)$
is different from Eq.~(\ref{eq: H to -=00005Cinfscaling}):
\begin{equation}
-\ln\mathcal{P}(H,L,t)\simeq\frac{H^{5/2}}{\sqrt{t}}\,f_{+}\left(\frac{L}{\sqrt{Ht}}\right).\label{eq: H to =00005Cinfscaling}
\end{equation}
In this limit one can neglect the diffusion term in Eq.~(\ref{eq:KPZ_dimensional}) \cite{KK2007,MKV}. The
resulting OFM equations describe a compressible flow of an effective
gas with negative pressure \cite{MKV}. For a finite $L$ these effective hydrodynamic equations
are still hard to solve analytically. Therefore, we evaluate the function
$f_{+}$ numerically, see Fig. \ref{fig: H>>1 s(L/sqrt(H))}. For this purpose we develop a special numerical
method, which employs Lagrangian coordinates and ultimately boils
down to solving a nonlinear second-order elliptic equation. Similarly to the function $f_{-}$, the function
$f_{+}$ describes a sharp transition from an $L$-dependent
solution to an $L$-independent one. This transition occurs at a critical
value $L_{\text{cr}}\simeq2\sqrt{2|H|t}/\pi$, which can be determined
analytically. By analogy with the ring problem \cite{SMS2018},
we believe that this transition has a fractional order $5/2$. It is
smoothed, however, by small diffusion effects. A schematic phase
diagram, showing different asymptotic behaviors of $\mathcal{P}(H,L,t)$
in the $\left(L/\sqrt{\nu t},\,\left|\lambda\right|H/\nu\right)$
plane, is shown in Fig. \ref{fig: phase diagram}.

The remainder of the paper is structured as follows. In Sec. \ref{Sec: OFM}
we briefly outline the OFM formulation of the problem. In Sec. \ref{Sec: lineartheory}
we address typical fluctuations of height, $\left|H\right|\ll\nu/\left|\lambda\right|$
and determine their dependence on $L$. Section \ref{Sec: negativetail}
deals with the negative tail of the height distribution. Here we employ some previously known exact static and moving soliton/ramp solutions to the OFM equation
to construct an approximate solution to the half-line problem at different
$L$. In this way we uncover a first-order dynamical phase transition
from an $L$-dependent ``phase" to an $L$-independent one. Sec. \ref{Sec: positivetail}
focuses on the opposite, positive tail of $\mathcal{P}(H,L,t)$. Here
we solve numerically an effective hydrodynamic problem. The solution
yields the optimal paths of the interface, the desired asymptotic
of the large deviation function of the height, and a dynamical
phase transition which, we believe, is of fractional order $5/2$.  We briefly summarize and discuss our results
in Sec. \ref{Sec: summary}. Some technical details are relegated to three Appendices.

\section{OFM formulation}
\label{Sec: OFM}

Let  $T$  be the measurement time of the interface height at $x=L$: $H=h\left(L,T\right)$. It is convenient to
write Eq. \eqref{eq:KPZ_dimensional} in a dimensionless form using
the scaling transformation $t/T\to t$, $x/\sqrt{\nu T}\to x$, and
$\left|\lambda\right|h/\nu\to h$:
\begin{equation}
\partial_{t}h
=\partial_{x}^{2}h
-\frac{1}{2}\left(\partial_{x}h\right)^{2}
+\sqrt{\epsilon}\,\xi\left(x,t\right),\label{eq: dimesionless KPZ}
\end{equation}
where $\epsilon= D\left|\lambda\right|^{2}\sqrt{T}/\nu^{\nicefrac{5}{2}}$
is the rescaled noise magnitude. The rescaled measurement coordinate
is
\begin{equation}
\ell=L/\sqrt{\nu T}, \label{eq: ell definition}
\end{equation}
and we are interested in the rescaled probability distribution
$\mathcal{P}\left(H,\ell\right)$. In the short-time limit, $\epsilon\to0$,
the exact path integral, corresponding to Eq. \eqref{eq: dimesionless KPZ},
can be evaluated using Laplace's method. This procedure boils down
to a minimization problem for the action
\begin{equation}
s\left[h\left(x,t\right)\right]
=\frac{1}{2}\int_{0}^{1}dt\int_{0}^{\infty}dx\left[\partial_{t}h
-\partial_{x}^{2}h+\frac{1}{2}\left(\partial_{x}h\right)^{2}\right]^{2},\label{eq: action}
\end{equation}
We define the Lagrangian
$$
\mathscr{L}\left[h(x,t)\right]=
\frac{1}{2}\int_{0}^{\infty}dx\left[\partial_{t}h-\partial_{x}^{2}h+\frac{1}{2}\left(\partial_{x}h\right)^{2}\right]^{2}
$$
such that $s=\int_{0}^{1}\mathscr{L}dt$, and introduce the conjugate
momentum via the variational derivative $\rho= \delta\mathscr{L}/\delta\left(\partial_{t}h\right)$.
The optimal path, in terms of $h(x,t)$ and $\rho(x,t)$,
solves the equations
\begin{align}
\partial_{t}h & =\partial_{x}^{2}h-\frac{1}{2}\left(\partial_{x}h\right)^{2}+\rho,\label{eq: OFM eq for h}\\
\partial_{t}\rho & =-\partial_{x}^{2}\rho-\partial_{x}\left(\rho\partial_{x}h\right),\label{eq: OFM eq for rho}
\end{align}
conditioned on $H=h\left(\ell,1\right)$. Comparing Eqs.~\eqref{eq: OFM eq for rho} and \eqref{eq: dimesionless KPZ}, we see that the conjugate momentum $\rho(x,t)$ -- a deterministic field -- describes the optimal realization of the noise $\sqrt{\epsilon} \xi(x,t)$.

The condition $H=h\left(\ell,1\right)$ can be accounted
for by introducing a Lagrange multi plier $\Lambda$ to the action
functional, and it leads to a condition on
$\rho(x,t=1)$ \cite{KK2007,MKV}
\begin{equation}
\rho\left(x,t=1\right)=\Lambda\delta\left(x-\ell\right).\label{eq: rho final time condition}
\end{equation}
The flat initial condition is
\begin{equation}
h\left(x,t=0\right)=0,\label{eq: h initial condition}
\end{equation}
and the reflecting boundary condition at $x=0$ is given by
\begin{equation}
\partial_{x}h\left(x=0,t\right)=\partial_{x}\rho\left(x=0,t\right)=0.\label{eq: reflecting condition}
\end{equation}
The zero-flux condition on $\rho$ ensures that the boundary term
at $x=0$, coming from the integration by parts of the linear variation
of the action, vanishes as it should. In terms of $\rho$ -- the optimal realization of the noise filed -- the action~(\ref{eq: action})
is given by
\begin{equation}
s=\frac{1}{2}\int_{0}^{1}dt\int_{0}^{\infty}dx\,\rho^{2}\left(x,t\right),\label{eq: s in terms of rho}
\end{equation}
so we expect $\rho\left(x\to\infty,t\right)=0$ for the action to be finite.

Similarly to the previous works
\citep{KK2007,KK2008,KK2009,MKV,KMSparabola,Janas2016,MSchmidt2017,SMS2018,SKM2018,MV2018},
once the OFM problem is solved and the action~(\ref{eq: s in terms of rho}) is evaluated,
$\mathcal{P}\left(H,\ell\right)$
is given, in the leading order, by
\begin{equation}
-\ln\mathcal{P}\left(H,\ell\right)\simeq\frac{s\left(H,\ell\right)}{\epsilon}.\label{eq: -lnP =00003D s/epsilon}
\end{equation}
Back in the physical (dimensional) variables,
we arrive at the scaling behavior~\eqref{eq: physical variables P(H,L,t)}.

\section{Typical fluctuations}

\label{Sec: lineartheory}For sufficiently small $\left|H\right|$,
that is, typical height fluctuations, the OFM problem can be solved
using a regular perturbation theory in $\left|H\right|$
or $\Lambda$ \cite{MKV}. The leading order of
$\mathcal{P}\left(H,\ell\right)$ is obtained by dropping the nonlinear
terms in Eq. \eqref{eq: OFM eq for h} and \eqref{eq: OFM eq for rho}. This leads to
\begin{align}
\partial_{t}h & =\partial_{x}^{2}h+\rho,\label{eq: linear OFM h equation}\\
\partial_{t}\rho & =-\partial_{x}^{2}\rho.\label{eq: linear OFM rho equation}
\end{align}
These linear equations are the (rescaled) OFM equations for the
Edwards-Wilkinson equation \cite{EdwardsWilkinson}
\begin{equation}
\partial_{t}h=\nu\partial_{x}^{2}h+\sqrt{D}\xi\left(x,t\right).\label{eq: EW equation}
\end{equation}
The solution to the antidiffusion equation \eqref{eq: linear OFM rho equation} with
the initial condition~\eqref{eq: rho final time condition}
and the reflecting boundary condition~\eqref{eq: reflecting condition} is
\begin{equation}
\rho\left(x,t\right)=\frac{\Lambda}{\sqrt{4\pi\left(1-t\right)}}\left[e^{-\frac{\left(x-\ell\right)^{2}}{4\left(1-t\right)}}+e^{-\frac{\left(x+\ell\right)^{2}}{4\left(1-t\right)}}\right].\label{eq: linear rho solution}
\end{equation}
To calculate the action, we plug $\rho\left(x,t\right)$
in Eq.~\eqref{eq: s in terms of rho} and use the fact that the integrand, when extended to the whole line $|x|<\infty$,
is an even function of $x$. This yields
\begin{equation}
s=\frac{\Lambda^{2}}{4}\left[I\left(\ell,1,\ell\right)+2I\left(\ell,1,-\ell\right)+I\left(-\ell,1,-\ell\right)\right],\label{eq: linear S in terms of I}
\end{equation}
where $I\left(x,t,x_{0}\right)$ is the double integral
\begin{eqnarray}
  I\left(x,t,x_{0}\right) &=& \int_{0}^{t}\frac{ds}{4\pi\sqrt{\left(t-s\right)\left(1-s\right)}}\nonumber \\
  && \int_{-\infty}^{\infty}d\xi e^{-\frac{\left(\xi-x\right)^{2}}{4\left(t-s\right)}-\frac{\left(\xi-x_{0}\right)^{2}}{4\left(1-s\right)}}.\label{eq: integral definition}
\end{eqnarray}
We evaluate this
integral in Appendix \ref{Appendix: evaluating the integral} and find that
\begin{align}
I\left(x,t,x_{0}\right) & =\frac{x-x_{0}}{4\sqrt{\pi}}\left[f\left(\frac{x-x_{0}}{\sqrt{4\left(1+t\right)}}\right)\right.\nonumber \\
 & \left.-f\left(\frac{x-x_{0}}{\sqrt{4\left(1-t\right)}}\right)\right],\label{eq: integral solution}
\end{align}
where $f\left(z\right)=e^{-z^{2}}/z+\sqrt{\pi}\erf\left(z\right)$, and $\erf$ is the error function. As a result,
\begin{equation}
s\left(\Lambda,\ell\right)=\frac{\Lambda^{2}}{2\sqrt{2\pi}}\left[1
+e^{-\frac{\ell^{2}}{2}}-\sqrt{\frac{\pi}{2}}\ell\erfc\left(\frac{\ell}{\sqrt{2}}\right)\right].
\label{eq: linear S in terms of =00005CLambda}
\end{equation}
Here $\erfc\left(z\right)=1-\erf\left(z\right)$ is the complementary
error function. Finally, we use the universal relation
\begin{equation}
\frac{ds}{d\Lambda}=\Lambda\frac{dH}{d\Lambda}\label{eq: Naftali's relation}
\end{equation}
 to express $\Lambda$ via $H$:
\begin{equation}
\Lambda\left(H,\ell\right)=\frac{\sqrt{2\pi}\,H}{1+e^{-\frac{\ell^{2}}{2}}-\sqrt{\frac{\pi}{2}}\ell\erfc\left(\frac{\ell}{\sqrt{2}}\right)},\label{eq: linear =00005CLambda(H, =00005Cell)}
\end{equation}
and arrive at
\begin{equation}
s\left(\left|H\right|\ll1,\ell\right)=\frac{\sqrt{\pi/2}\,H^{2}}{1+e^{-\ell^{2}/2}-\sqrt{\pi/2}\,\ell\erfc\left(\ell/\sqrt{2}\right)}.\label{eq: linear s(H,=00005Cell)}
\end{equation}
As to be expected, the action is quadratic in $H$, so for typical
height fluctuations the one-point distribution $\mathcal{P}\left(H,\ell\right)$
is Gaussian in $H$.

\begin{figure}
\centering{}\includegraphics[scale=0.45]{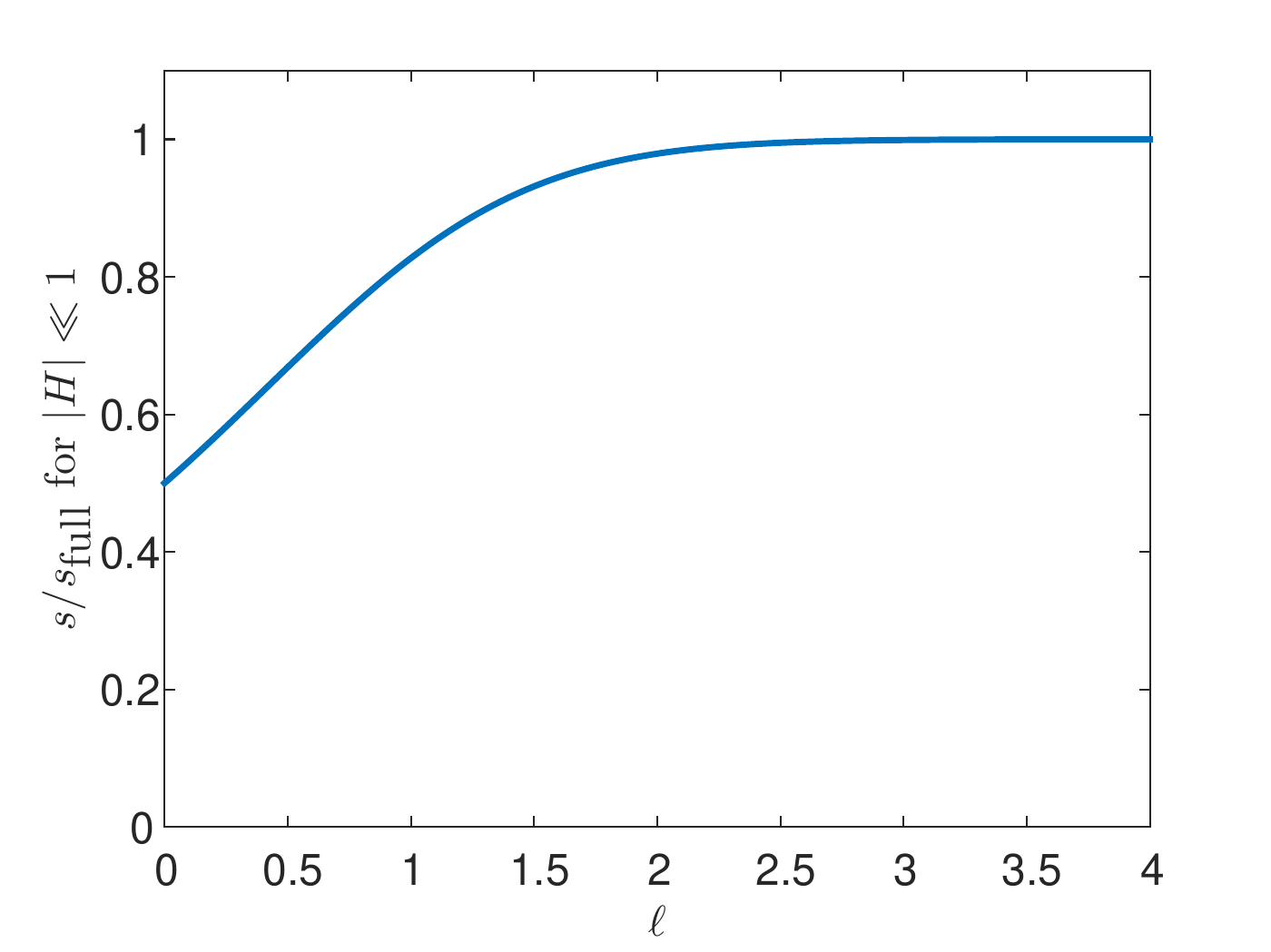}
\caption{The half-line action (in the units of the full-line
action) vs. $\ell$  for typical fluctuations.}
\label{fig: linear action}
\end{figure}

For the full-line system the action for typical fluctuations is $s_{\text{full}} =\sqrt{\pi/2}\,H^{2}$
\cite{MKV}, so the ratio $s/s_{\text{full}}$ depends only on $\ell$. This dependence is
shown in Fig. \ref{fig: linear action}. At $\ell=0$ we obtain  $s/s_{\text{full}}=1/2$, as to be expected from symmetry arguments. At small but nonzero $\ell$ we obtain a linear dependence
\begin{equation}
\frac{s(\ell)}{s_{\text{full}}}\simeq \frac{1}{2}+\frac{1}{4} \sqrt{\frac{\pi }{2}} \ell,\quad \ell\ll 1, \label{eq: linear dependence small ell}
\end{equation}
while at $\ell\to\infty$ $s(\ell)/s_{\text{full}}$  approaches $1$.

\begin{figure}
\includegraphics[scale=0.45]{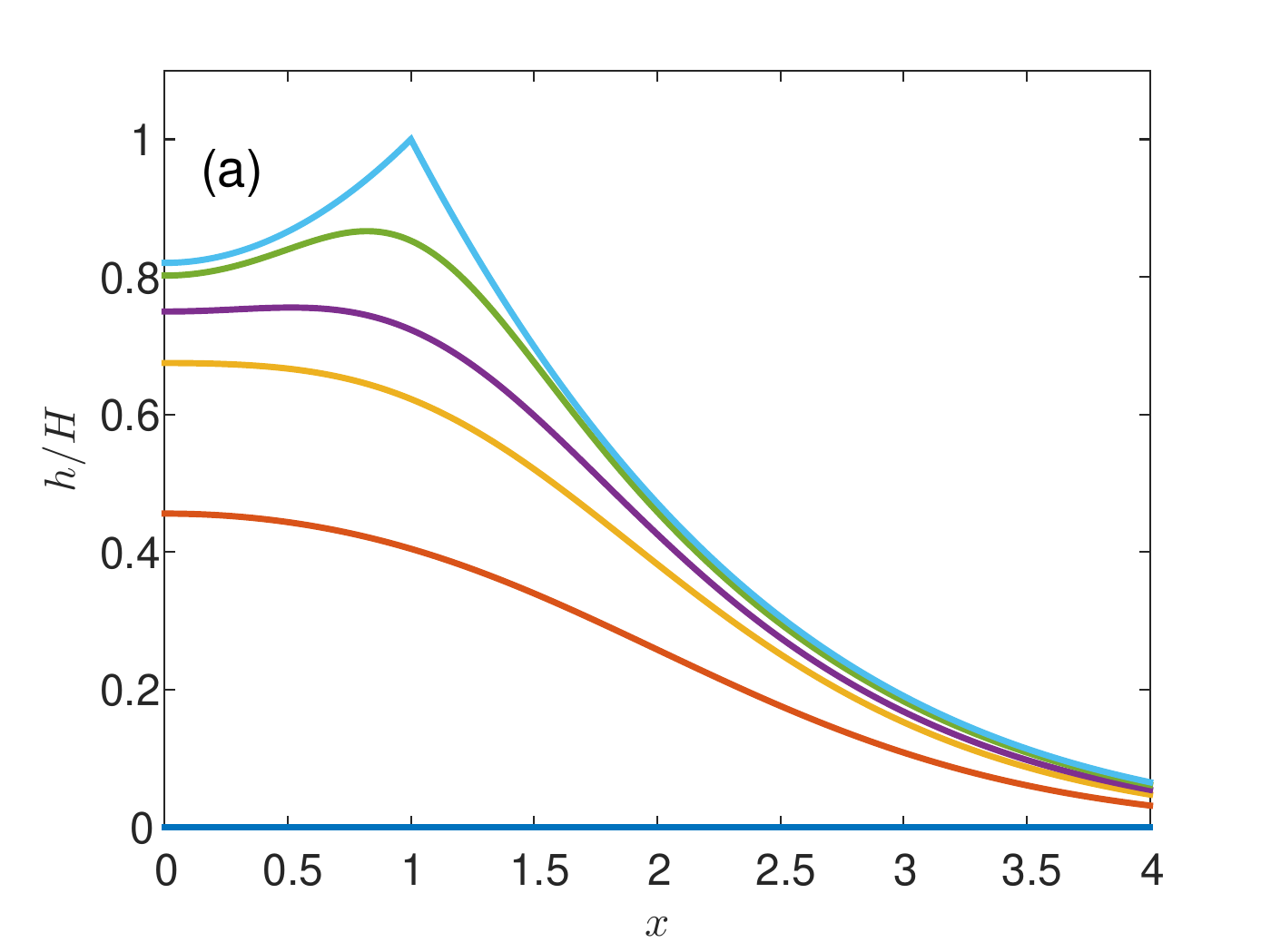}
\includegraphics[scale=0.45]{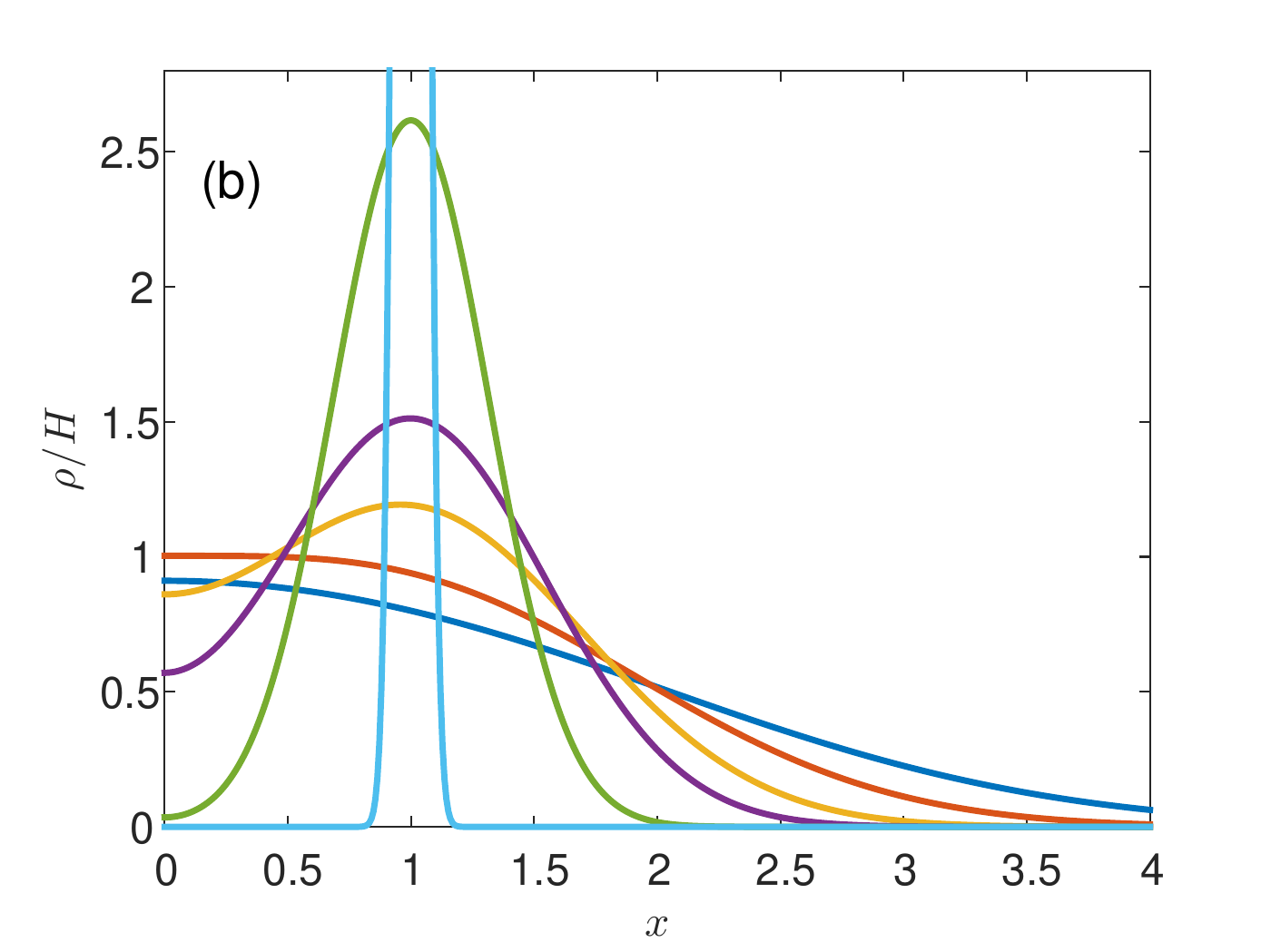}
\caption{The optimal path $h(x,t)/H$ (a) and $\rho(x,t)/H$ (b) as described by the linear theory
(Eqs. \eqref{eq: linear h solution} and \eqref{eq: linear rho solution}, respectively)  for $\ell=1$. The $x$-profiles are shown at rescaled times  $t=0$, $0.5$, $0.75$, $0.85$, $0.95$ and $t=1$ (from bottom to top) for $h$, and at the same times for $\rho$, except that $t=1$ is replaced by $t=0.999$. Notice the corner singularity of $h(x,t=1)$ at $x=\ell$.}
\label{fig: linear profiles}
\end{figure}

In order to determine the most probable height history, we should solve
Eq. \eqref{eq: linear OFM h equation}:
a diffusion equation with $\rho\left(x,t\right)$ acting as a source
term. Its solution for $x\ge0$ for the initial condition \eqref{eq: h initial condition}
and the reflecting boundary condition \eqref{eq: reflecting condition}
is given by
\begin{alignat}{1}
h\left(x,t\right) & =\int_{0}^{t}ds\int_{0}^{\infty}d\xi\left[G\left(x-\xi,t-s\right)\right.\nonumber \\
 & \left.+G\left(x+\xi,t-s\right)\right]\rho\left(\xi,s\right),\label{eq: linear h solution expression}
\end{alignat}
where $G\left(x,t\right)=e^{-\frac{x^{2}}{4t}}/\sqrt{4\pi t}$ is
the Green's function for the diffusion equation. Plugging here $\rho\left(x,t\right)$
from Eq. \eqref{eq: linear rho solution} we arrive at
\begin{equation}
h\left(x,t\right)=\Lambda\left(H,\ell\right)\left[I\left(x,t,\ell\right)+I\left(x,t,-\ell\right)\right]\label{eq: linear h solution}
\end{equation}
with $\Lambda\left(H,\ell\right)$ from Eq. \eqref{eq: linear =00005CLambda(H, =00005Cell)}
and $I\left(x,t,x_{0}\right)$ from Eq. \eqref{eq: integral solution}.
Figure \ref{fig: linear profiles} shows the rescaled optimal height history $h(x,t)/H$
and rescaled optimal noise realization for $\ell=1$.

\section{$\lambda H\to\infty$ tail}
\label{Sec: negativetail}

Now we consider the $H\to-\infty$ tail of $\mathcal{P}\left(H,\ell\right)$. Here, as well as in the
opposite tail $H\to\infty$, the optimal path of the system is dominated by the nonlinearity of the KPZ equation.
However, in  contrast to the $H\to\infty$ tail, the optimal realization of the noise $\rho(x,t)$ in this tail is
localized in a small region of space, so that one cannot neglect the diffusion term in the KPZ equation \cite{KK2007,MKV,KMSparabola,Janas2016,MSchmidt2017,SKM2018,MV2018}.  As we found, two exact particular soliton
solutions to Eqs.~\eqref{eq: OFM eq for h} and \eqref{eq: OFM eq for rho} serve as
``building blocks'' of the approximate solution to this problem, based on the large parameter $|H|$. These particular solutions have previously appeared in other settings \cite{Mikhailov1991,Fogedby1999,KK2007,MKV,Janas2016}.

The first exact particular solution is the \emph{static} soliton solution, which
involves a localized stationary $\rho$-profile, which
we call a soliton, and a vertically traveling $h$-profile \cite{Mikhailov1991,Fogedby1999,KK2007,MKV}:
\begin{align}
\rho\left(x\right) & =-2c\sech^{2}\left[\sqrt{c/2}\left(x-x_0\right)\right],\label{eq: static soliton rho}\\
h\left(x,t\right) & =2\ln\left\{ \cosh\left[\sqrt{c/2}\left(x-x_0\right)\right]\right\} -ct,\label{eq: static soliton h}
\end{align}
with a constant $x_0$ and a constant $c>0$.  The second exact particular solution is the \emph{traveling}
soliton solution, where a $\rho$-soliton travels along
the $x$ axis without changing its shape, and $h$ behaves as a traveling
``ramp''. For the right moving soliton the profiles are given by
\cite{Mikhailov1991,Fogedby1999,Janas2016}
\begin{align}
\rho\left(x,t\right) & =-c^{2}\sech^{2}\left[\frac{c}{2}\left(x_{0}-x+ct\right)\right],\label{eq: traveling solition rho}\\
h\left(x,t\right) & =2\ln\left[1+e^{c\left(x_{0}-x+ct\right)}\right]-2c\left(ct-x\right).\label{eq: traveling soliton h}
\end{align}
Here the soliton is centered at $x=x_{0}+ct$, where the constant
$c>0$ is the soliton speed.
A left-moving soliton can be obtained by replacing $c$ by $-c$.

As we will show now,  when $c\gg 1$, the first of these two exact solutions, and a nontrivial combination of the first and second solutions, can be used, alongside with the trivial solution $\rho=h=0$,
to approximately satisfy (up to small boundary layers and transients) the
boundary conditions~\eqref{eq: rho final time condition}-\eqref{eq: reflecting condition}.
The two resulting solutions, which we call static and dynamic {[}because
of the behavior of their $\rho(x,t)${]}, yield different actions, leading to a first-order dynamical phase transition.

\subsection{Static solution}
\label{subsec: Static-solution}

\begin{figure}
\includegraphics[scale=0.5]{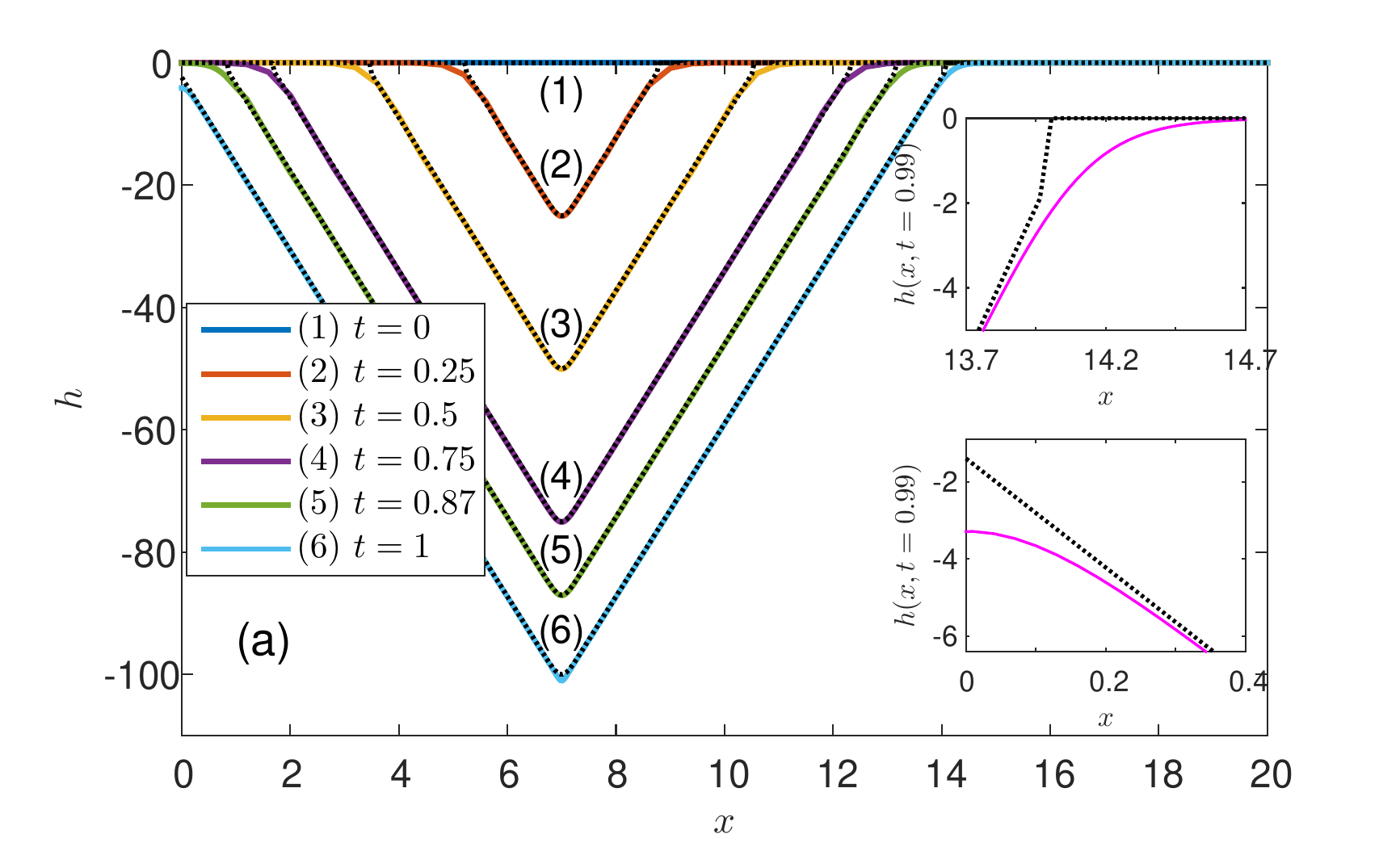}
\includegraphics[scale=0.5]{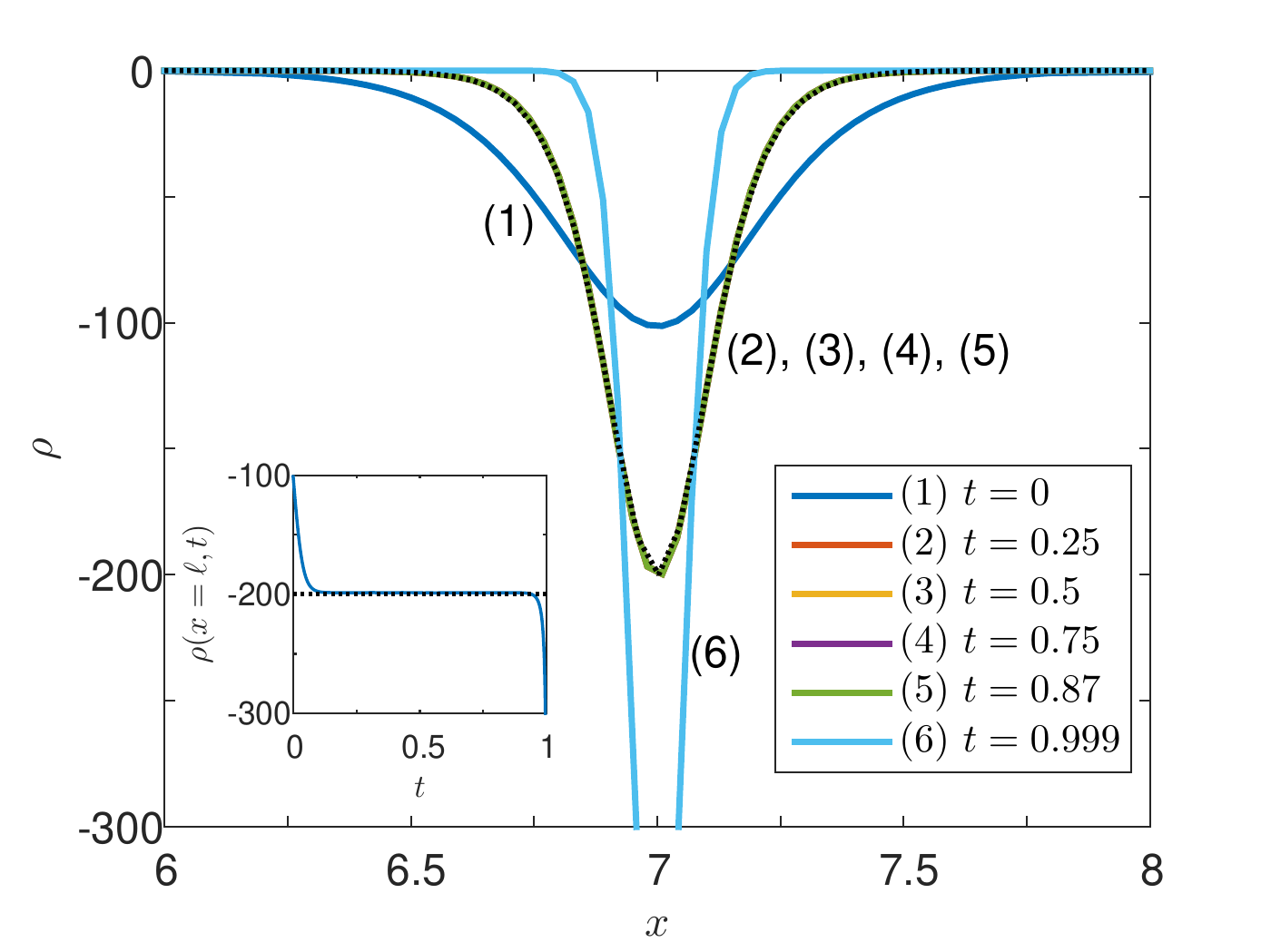}
\caption{The optimal path $h(x,t)$ (a) and $\rho(x,t)$ (b), described by the static solution for the $\lambda H>0$ tail. The parameters are $H=-100$
and $\ell=7$. Shown are numerical results (solid lines) and  analytical predictions of Eqs.~\eqref{eq: static soliton rho}
and~\eqref{eq: static soliton h} (dotted lines) at indicated times.
Upper inset of (a): $h\left(x,t=0.99\right)$
inside the boundary layer at $x=\ell+\sqrt{c/2}\,t$. Lower inset of (a): $h\left(x,t=0.99\right)$
inside the boundary layer at $x=0$. Inset of (b): $\rho\left(x=\ell,t\right)$.
Clearly, $\rho\left(x,t\right)$ does not change in time except during
narrow transients near $t=0$ and $t=1$. The numerical
and analytical curves are only distinguishable in the insets in (a)
and during  the short transients in (b). The numerical solution captures the boundary layers of the $h$-profile, unaccounted for by Eq. (\ref{eq: traveling soliton h}).}
\label{fig: negative tail static profiles}
\end{figure}

The static solution is described by Eqs.~\eqref{eq: static soliton rho} and
\eqref{eq: static soliton h} with $x_0=\ell$. This solution, see Fig. \ref{fig: negative tail static profiles}, is very similar to the solution, which determines the  $\lambda H>0$ tail of  the full-line problem \cite{KK2007,MKV}. It immediately follows from Eq.~(\ref{eq: static soliton h}) and the condition $h(0,1)=H$ that we must set $c=-H\gg 1$. As in Refs. \cite{KK2007,MKV},  Eq.~(\ref{eq: static soliton rho}) does not satisfy the final-time condition~\eqref{eq: rho final time condition}. The exact solution to the problem develops a short transient close to $t=1$, which takes care of this boundary condition, similarly to Ref. \cite{MKV}. Another short transient appears
close to $t=0$, see the inset in Fig. \ref{fig: negative tail static profiles}(b). The contributions of these
transients to the action are of a subleading order in
$\left|H\right|\gg1$ and, similarly to Refs. \cite{KK2007,MKV}, we will ignore them.

Equations~\eqref{eq: static soliton rho} and \eqref{eq: static soliton h}
apply only on a finite interval $x_{1}(t)<x<x_{2}(t)$, where
$x_{2}(t)=\ell+\sqrt{c/2}\,t$ as in the full-line problem
\cite{MKV}, whereas
\begin{equation}
x_{1}\left(t\right)=\begin{cases}
\ell-\sqrt{c/2}\,t, & 0<t<\frac{\ell\sqrt{2}}{\sqrt{c}},\\
0, & \frac{\ell\sqrt{2}}{\sqrt{c}}<t<1 .
\end{cases}\label{x2}
\end{equation}
At $x>x_{2}(t)$, and at $0<x<x_{1}(t)$ and $0<t<\ell\sqrt{2}/\sqrt{c}$
one can use the trivial solution $\rho=h=0$, see Fig. \ref{fig: negative tail static profiles}. There are two boundary layers, at $x_1(t)$ and $x_2(t)$, but they only give subleading corrections to the action. As was shown in Ref. \cite{MKV}, the moving boundary layer at $x_2(t)$ is a shock of the Burgers equation
\begin{equation}\label{Burgers}
\partial_tV+V\partial_xV=\nu \partial_x^2V
\end{equation}
or, if one neglects the diffusion term, of the Hopf equation
\begin{equation}
\partial_{t}V+V\partial_{x}V=0 \label{eq: Hopf equation}
\end{equation}
for the interface slope
\begin{equation}
V\left(x,t\right)=\partial_{x}h\left(x,t\right). \label{eq: V definition}
\end{equation}

The characteristic soliton width is $w\sim 1/\sqrt{c}=1/\sqrt{|H|}$. At $\left|x-\ell\right|\gg w$,
$\rho(x)$ decays exponentially. As a result, the reflecting boundary condition \eqref{eq: reflecting condition}
for $\rho$ is satisfied up to exponentially small corrections provided
that $\ell\gg1/\sqrt{\left|H\right|}$, that is $\left|H\right|\gg 1/\ell^2$.

We verified the static solution by solving the full OFM problem, formulated in Sec.~\ref{Sec: OFM},  numerically. As in the previous works  \citep{MKV,Janas2016,SMS2018,SKM2018,MV2018}, we used the Chernykh-Stepanov back-and-forth iteration algorithm \cite{ChernykhStepanov}. Here we started the iteration procedure sufficiently close to the expected solution. A comparison of the analytic and numerical results for the static solution is presented in Fig.~\ref{fig: negative tail static profiles}.

Because of the strong localization of the $\rho$-soliton, the
rescaled action \eqref{eq: s in terms of rho} of this solution does not depend on $\ell$ and coincides with the
corresponding expression for the full-line system \cite{KK2007,MKV}:
\begin{equation}
s_{\text{s}}\left(H\to-\infty,\ell\right)\simeq\frac{8\sqrt{2}}{3}\left|H\right|^{3/2},\label{eq: static solution action}
\end{equation}
where the subscript s stands for ``static".

\subsection{Dynamic solution}
\label{subsec: Dynamic-solution}

\begin{figure}
\includegraphics[scale=0.5]{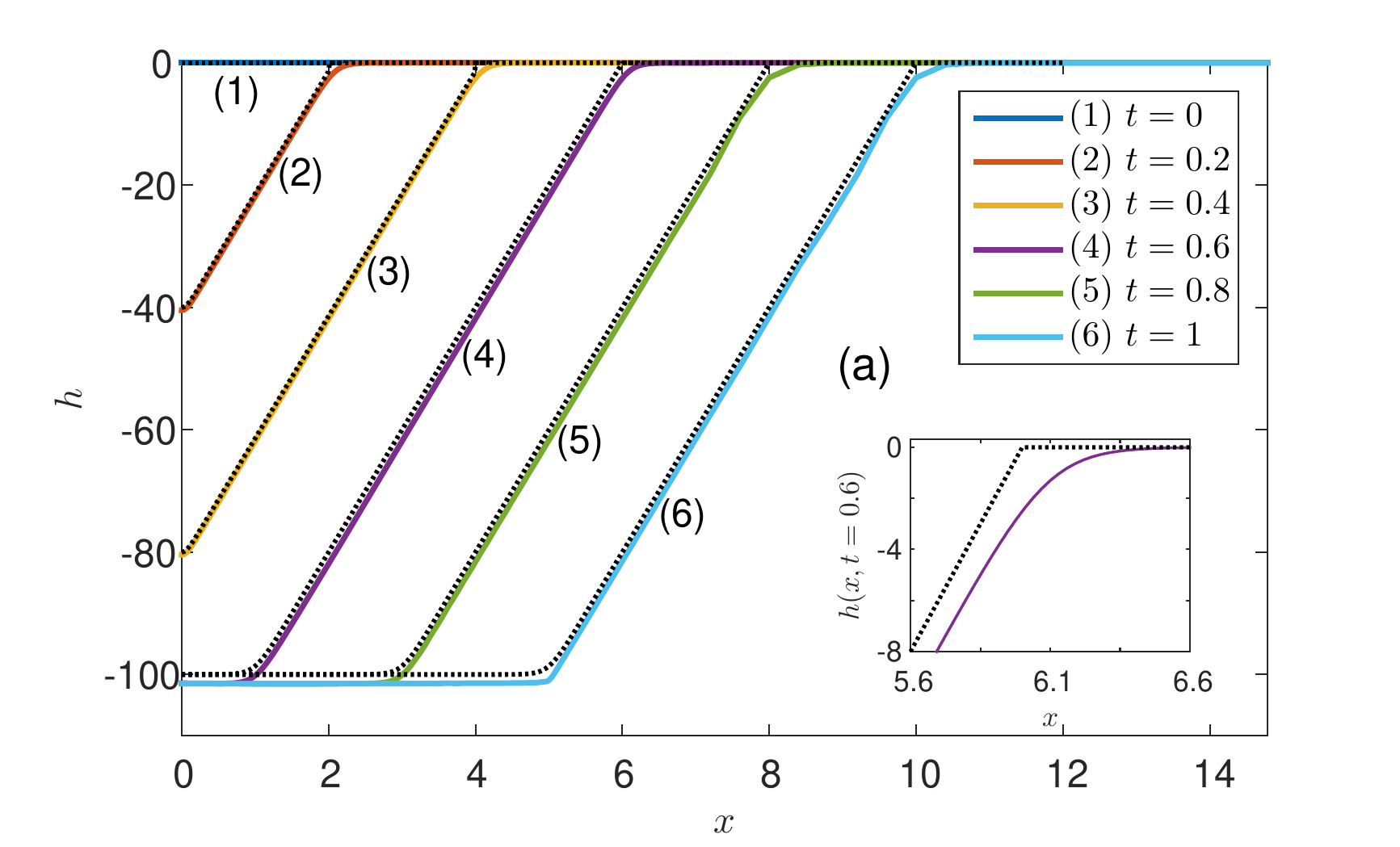}
\includegraphics[scale=0.5]{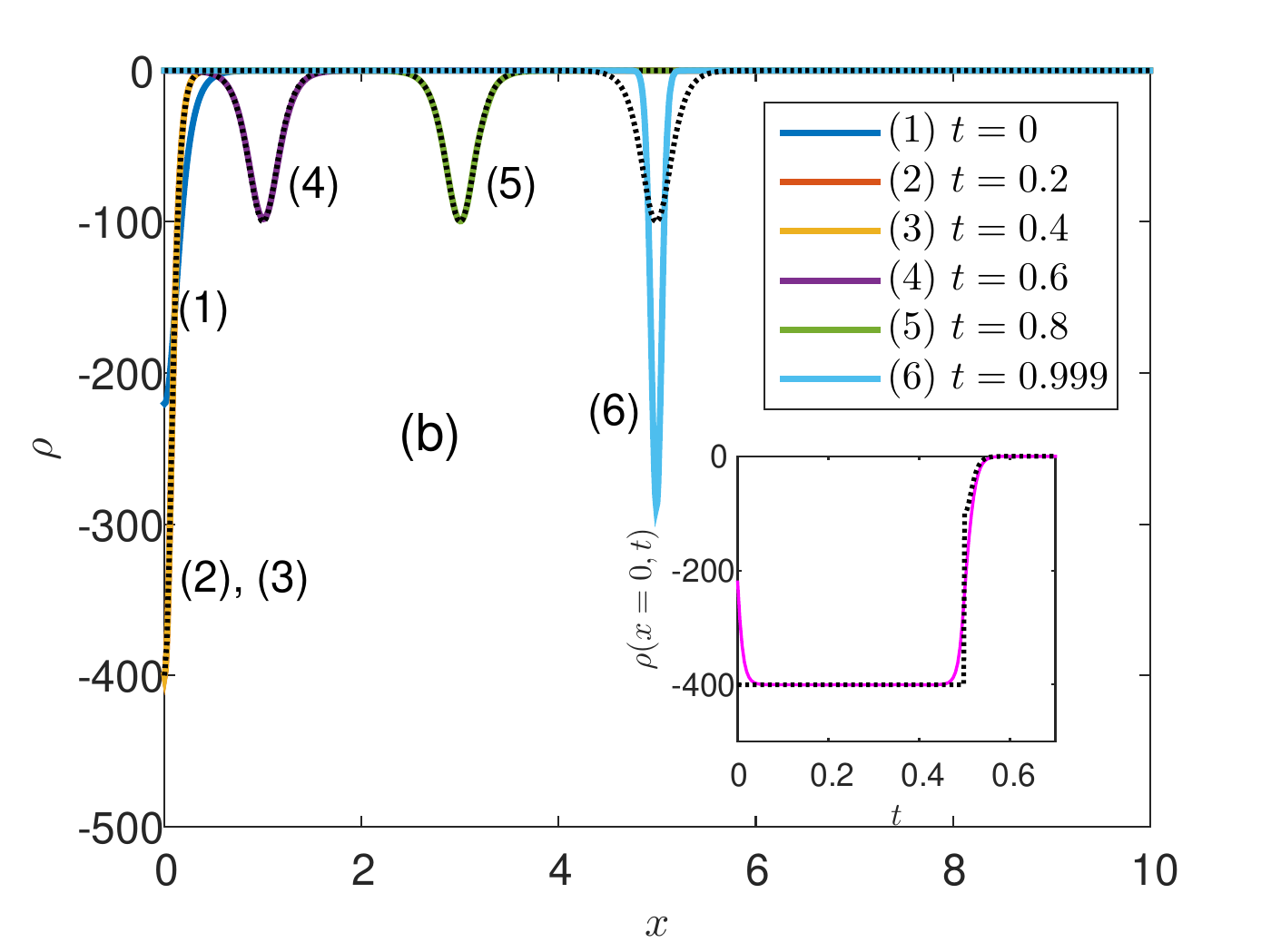}
\caption{The optimal path $h(x,t)$ (a) and $\rho(x,t)$ (b), described by the dynamic solution for the $\lambda H>0$ tail. The parameters are $H=-100$
and $\ell=5$, for which $c_{2}\simeq10.68$ and $\tau\simeq0.44$, see the main text. Shown are numerical results (solid lines) and
analytical predictions from Eqs.~\eqref{eq: dynamic solution h}
and \eqref{eq: dynamic solution rho} (dotted lines) at indicated times.
Inset of (a) shows the boundary layer of $h$ at $x=c_{2}t$, captured by the numerical solution at $t=0.6$. Inset of
(b) shows the short transients at $t=0$ and $t=\tau$, captured by the numerical solution for $\rho\left(x=0,t\right)$. The $\rho$-soliton changes
its amplitude in accordance with Eq.~(\ref{eq: rho conservation law}) as it changes from the static soliton to the traveling one.}
\label{fig: negaive tail dynamic profiles}
\end{figure}

The dynamic solution involves (quite a fascinating)  metamorphosis between the static and traveling soliton solutions. At very short times the static soliton solution Eqs.~\eqref{eq: static soliton rho} and
\eqref{eq: static soliton h} is formed at $x=0$ and persists until some intermediate time $0<\tau<1$. Then the static soliton solution rapidly turns into a traveling soliton/ramp solution of the type~(\ref{eq: static soliton rho}) and~(\ref{eq: static soliton h}). The latter moves to the right and reaches the point $x=\ell$ at time very close to $1$, where $\rho$ rapidly becomes delta-function (see Fig.~\ref{fig: negaive tail dynamic profiles}).
In the region where this solution predicts $h>0$, we should use the trivial solution $h=0$.   Why is such a surprisingly complex solution possible?

To begin with, by virtue of the reflecting boundary condition at $x=0$, our half-line problem is equivalent to the right half, $x\geq 0$, of a symmetric full-line
problem where the dynamics of initially flat KPZ interface is conditioned on reaching the height $H$ at time $1$ at \emph{two} symmetric points $x=\ell$ and $x=-\ell$.  It was previously shown
that the OFM equations~(\ref{eq: OFM eq for h}) and~(\ref{eq: OFM eq for rho}) have two families of exact multi-soliton solutions \cite{Janas2016}. Among them there is a solution where a single $\rho$-soliton stays at $x=0$ (and drives a vertically traveling $h$-front) until some time $t=\tau$, and then splits into two outgoing traveling solitons (which drive two outgoing $h$-ramps). For large $c$ the splitting process is vert short. As a result, for most of the time, this exact solution can be approximated as a time sequence of two simpler solutions: a solution describing a static $\rho$-soliton at $x=0$, and a solution describing two individual $\rho$-solitons, traveling to the right and to the left, respectively, and driving two outgoing $h$-ramps.  The splitting time $\tau$ of the static soliton can be anywhere  between
$t=0$ and $t=1$ depending on $c$ and on other constants \cite{Janas2016}. The $x>0$ part of this solution is what we call the dynamic solution to our half-line problem.  We present this solution in Appendix \ref{Appendix: dynamic from multisoliton}. In the full solution of the problem the traveling soliton reaches $x=\ell$ at $t$ very close to $1$ where it rapidly becomes the delta function.

Using Eqs. \eqref{eq: static soliton rho} and \eqref{eq: traveling solition rho}
and the fact that the traveling soliton must be located at $x=0$
at $t=\tau$, we can write the $\rho$-profile of the dynamic solution
as
\begin{equation}
\rho\left(x,t\right)\simeq\begin{cases}
-2c_{1}\sech^{2}\left(\sqrt{c_{1}/2}\,x\right),\! \!\!& 0<t<\tau,\\
-c_{2}^{2}\sech^{2}\left\{ \frac{c_{2}}{2}\left[-x+c_{2}\left(t-\tau\right)\right]\right\} , \!\!\!& \tau<t<1,
\end{cases}\label{eq: dynamic solution rho c1 c2}
\end{equation}
One relation between the soliton parameters $c_{1}$ and $c_{2}$ can be found from the conservation law
\begin{equation}
\int_{0}^{\infty}\rho\left(x,t\right)dx=\text{const}, \label{eq: rho conservation law}
\end{equation}
which immediately follows from Eq. \eqref{eq: OFM eq for rho} and the reflecting boundary conditions \eqref{eq: reflecting condition}.  The conservation law yields
\begin{equation}
c_{1}=2c_{2}^{2},\label{eq: c1=00003D2c2^2}
\end{equation}
and we will ultimately express $c_2$  via $H$ and $\ell$. We use the trivial solution $\rho=h=0$
at $x>c_{2}t$, where the traveling $h$-front~\eqref{eq: static soliton h},
and the traveling $h$-ramp \eqref{eq: traveling soliton h} are positive,
and ignore the boundary layers which smooth the transition between the nontrivial and trivial solutions.
Altogether, the dynamic solution is given by
\begin{widetext}
\begin{align}
\rho\left(x,t\right) & \simeq\begin{cases}
-4c_{2}^{2}\sech^{2}\left(c_{2}x\right), & 0<t<\tau,\\
-c_{2}^{2}\sech^{2}\left\{ \frac{c_{2}}{2}\left[-x+c_{2}\left(t-\tau\right)\right]\right\} , & \tau<t<1,
\end{cases}\label{eq: dynamic solution rho}\\
h\left(x,t\right) & \simeq\begin{cases}
2\ln\left[\cosh\left(c_{2}x\right)\right]-2c_{2}^{2}t, & 0<x<c_{2}t,\,0<t<\tau,\\
2\ln\left\{ 1+e^{c_{2}\left[-x+c_{2}\left(t-\tau\right)\right]}\right\} -2c_{2}\left(c_{2}t-x\right), & 0<x<c_{2}t,\,\tau<t<1,\\
0, & x>c_{2}t.
\end{cases}\label{eq: dynamic solution h}
\end{align}
\end{widetext}
In terms of the interface slope $V(x,t)=\partial_x h(x,t)$ the solution (\ref{eq: dynamic solution h}) for $\tau<t<1$ describes a shock-antishock pair, which propagates to the right with a constant speed $c_2$ \cite{Fogedby1998,Fogedby1999}. The two nontrivial expressions for $h$ in Eq.~(\ref{eq: dynamic solution h}) match at $t=\tau$ outside of the narrow transition
region between the static and traveling solitons:
\begin{align}
h\left(x\gg1/c_{2},t\to\tau^{-}\right) & \simeq h\left(x\gg1/c_{2},t\to\tau^{+}\right)\nonumber \\
& \simeq 2c_{2}x-2c_{2}^{2}\tau.\label{eq: dynamic h asymptotic match}
\end{align}
The flat initial condition \eqref{eq: h initial condition} is satisfied.  The reflecting boundary condition \eqref{eq: reflecting condition}
is satisfied both for $t<\tau$ and (up to exponentially small corrections) at $t>\tau$.  There are three short transients, unaccounted for by the
dynamic solution~(\ref{eq: dynamic solution rho}) and~(\ref{eq: dynamic solution h}): the first close to $t=0$, where the static soliton forms,
the second around $t=\tau$, where the static soliton becomes the traveling
one, and the third close to $t=1$, where the traveling soliton becomes
delta function. These transients do not contribute to the action in the leading order that we are after.

In order to express $c_{2}$ and $\tau$ through the parameters
$H$ and $\ell$, we employ the height condition $2c_2^2 \tau = |H|$ and the kinematic relation $c_2(1-\tau)=\ell$. These yield
\begin{align}
c_{2} & =\frac{1}{2}\left(\ell+\sqrt{\ell^{2}+2\left|H\right|}\right),\label{eq: c2(H,l)}\\
\tau & =\frac{|H|+\ell^2-\ell\sqrt{\ell^{2}+2\left|H\right|}}{|H|}. \label{eq: tau(H,l)}
\end{align}
We verified the dynamic solution numerically, see Fig. \ref{fig: negaive tail dynamic profiles}, by starting the Chernykh-Stepanov iteration procedure \cite{ChernykhStepanov} sufficiently closely to the expected solution.

Now we are in a position to evaluate the action~\eqref{eq: s in terms of rho}  for the dynamic solution. We use Eq.~(\ref{eq: dynamic solution rho}) and split the integration in time into two regions, $0<t<\tau$ and $\tau<t<1$:
\begin{align}
s & \simeq8c_{2}^{4}\int_{0}^{\tau}dt\int_{0}^{\infty}dx\sech^{4}\left(c_{2}x\right)\nonumber \\
 & +\frac{c_{2}^{4}}{2}\int_{\tau}^{1}dt\int_{0}^{\infty}dx\sech^{4}\left\{ \frac{c_{2}}{2}\left[-x+c_{2}\left(t-\tau\right)\right]\right\} .\label{eq: dynamic action computation}
\end{align}
Using Eqs.~(\ref{eq: c2(H,l)}) and (\ref{eq: tau(H,l)}),  we finally arrive
at
\begin{equation}
s_{\text{d}}\left(H\to-\infty,\ell\right)\simeq2\ell\left|H\right|+\frac{2}{3}\ell^{3}
+\frac{2}{3}\left(\ell^{2}+2\left|H\right|\right)^{3/2},\label{eq: dynamic solution action}
\end{equation}
where the subscript d stands for ``dynamic".

\subsection{Dynamical phase transition}
\label{subsec: Dynamical-phase-transition}

\begin{figure}
\centering{}\includegraphics[scale=0.5]{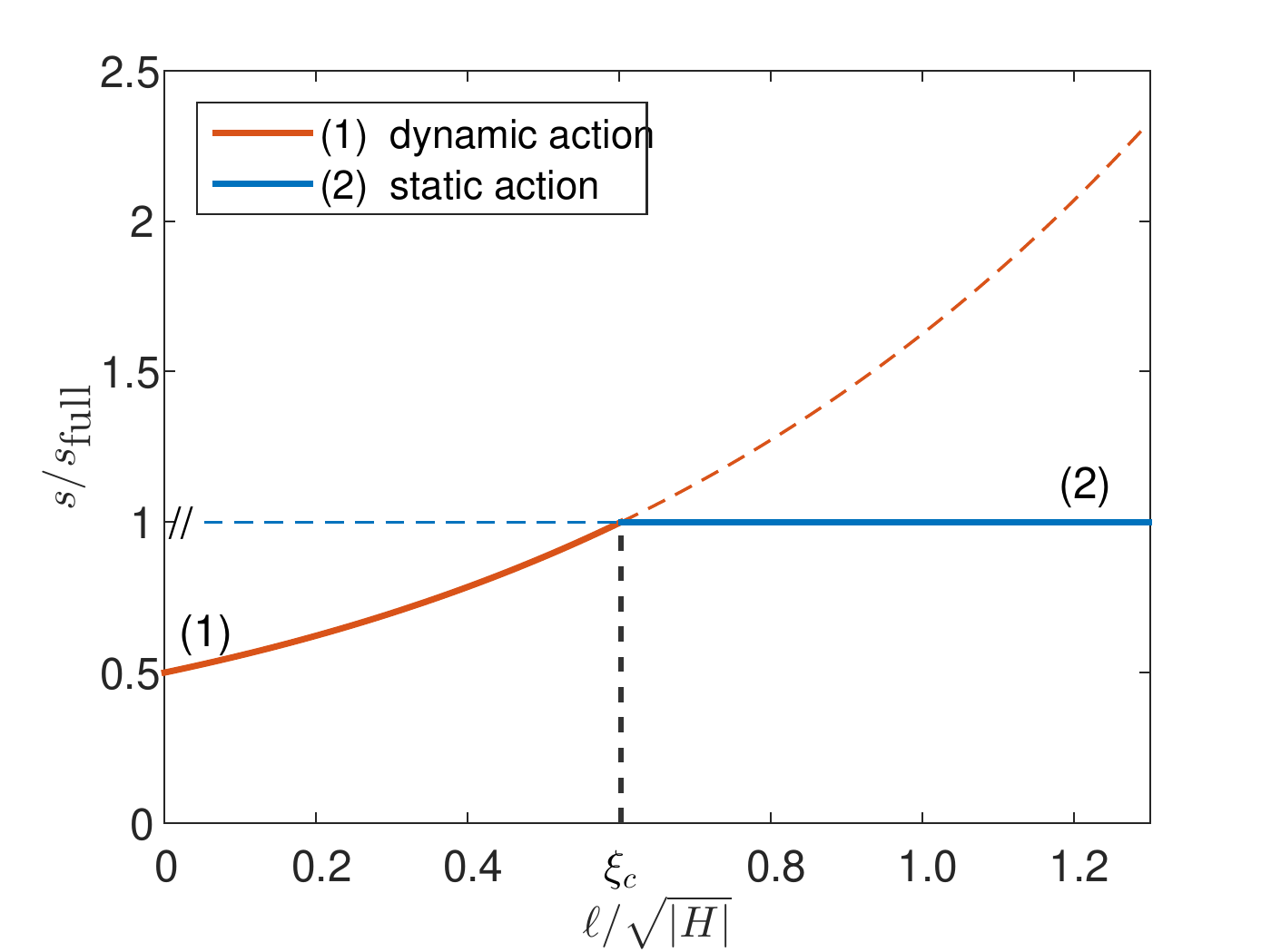}
\caption{The action $s(H\to-\infty,\ell)$ in the units of the full-line
action for $H\to-\infty$,  $s_{\text{full}}=8\sqrt{2}\left|H\right|^{3/2}/3$. The solid line indicates the least action for any
$\ell/\sqrt{\left|H\right|}$. The dashed lines are the
static and dynamic actions, respectively (see the main text), in the regions where they
are not minimal. The static action for very small $\ell/\sqrt{\left|H\right|}$ is not displayed,
since the static solution is invalid for $\ell\sim1/\sqrt{\left|H\right|}$.
Evident is a first-order dynamical phase transition at $\ell/\sqrt{|H|}=\xi_{c}=0.602239\dots$.}
\label{fig: negative tail action}
\end{figure}

When $|H|\gg \text{max}\, (1,1/\ell^2)$, each of the two solutions, the static and dynamic, exists for any $\ell>0$.
Their actions~(\ref{eq: static solution action}) and~(\ref{eq: dynamic solution action}) have a common factor $H^{3/2}$. In order to find the minimum action at specified $-H\gg 1$ and $\ell$, we can compare the quantities
\begin{align}
f_{\text{s}} & =\frac{s_{s}\left(H,\ell\right)}{\left|H\right|^{3/2}}\simeq\frac{8\sqrt{2}}{3},\label{eq: divided static action}\\
f_{\text{d}} & =\frac{s_{d}\left(H,\ell\right)}{\left|H\right|^{3/2}}\simeq \frac{2\ell}{\sqrt{\left|H\right|}}+\frac{2}{3}\left(\frac{\ell}{\sqrt{\left|H\right|}}\right)^{3}\nonumber \\
 &
 +\frac{2}{3}\left[\left(\frac{\ell}{\sqrt{\left|H\right|}}\right)^{2}+2\right]^{3/2}\label{eq: divided dynamic action}
\end{align}
These quantities are functions of the single variable $\xi=\ell/\sqrt{\left|H\right|}$,
and they are depicted in Fig. \ref{fig: negative tail action}.
As one can see, the dynamic solution is optimal for $\ell<\xi_c\sqrt{|H|}$, whereas
the static solution is optimal  for $\ell>\xi_c\sqrt{|H|}$. Here $\xi_c=0.602239\dots$ is
the root of the algebraic equation
$$
2 \xi+\frac{2}{3} \xi^3 +\frac{2}{3}\left(\xi^2+2\right)^{3/2}=\frac{8\sqrt{2}}{3}.
$$

Overall, the action is given by the scaling relation
\begin{equation}
s\left(H\to-\infty,\ell\right)
\simeq\left|H\right|^{3/2}f_{-}\left(\frac{\ell}{\sqrt{\left|H\right|}}\right),
\label{eq: action scaling negative tail}
\end{equation}
where
\begin{equation}
f_{-}\left(\xi\right)=\begin{cases}
2\xi+\frac{2}{3}\xi^{3}+\frac{2}{3}\left(\xi^{2}+2\right)^{3/2}, & \xi\le\xi_{c},\\
\frac{8\sqrt{2}}{3}, & \xi\ge\xi_{c}.
\end{cases}\label{eq: negative tail scaling funcion f_}
\end{equation}
This result leads to Eq.~(\ref{eq: H to -=00005Cinfscaling}), announced in the Introduction.
The first derivative of $s\left(H,\ell\right)$ with respect to $H$ is discontinuous, at large $-H$, across the parabola
$H=-\ell^2/\xi_c^2$ in the $\ell,H$ plane. Such singularities of the action are classified as first-order dynamical phase transitions.
In the limit of $\ell \to \infty$  the action  coincides with the expression $s_{\text{full}}=8\sqrt{2}\left|H\right|^{3/2}/3$ for the infinite line, obtained in Refs. \cite{KK2007,MKV}. In the limit of $\ell \to 0$ the action is given by Eq.~(\ref{eq: relation}).  That the switch between the two limits is observed at a finite $\ell$, via a first-order phase transition, is both interesting and unexpected.

\section{$\lambda H\to-\infty$ tail}
\label{Sec: positivetail}

The opposite tail, $H\to\infty$, is very
different in its nature. Here, as in the previous works \cite{KK2007,KK2009,MKV,KMSparabola,Janas2016,SKM2018,SMS2018,MV2018}, we can neglect the diffusion terms in Eqs.~\eqref{eq: OFM eq for h} and
\eqref{eq: OFM eq for rho}. Then, differentiating Eq. \eqref{eq: OFM eq for h}
with respect to $x$,
we arrive at  the equations
\begin{align}
\partial_{t}\rho & +\partial_{x}\left(\rho V\right)=0,\label{eq: continutity equation}\\
\partial_{t}V & +V\partial_{x}V=\partial_{x}\rho.\label{eq: momentum equation}
\end{align}
These equations, with the initial condition
\begin{equation}
V\left(x,t=0\right)=0\label{eq: V initial condition}
\end{equation}
and the final-time condition \eqref{eq: rho final time condition},
describe collapse of an initially static cloud of an inviscid
gas with density $\rho(x,t)$ and velocity
$V(x,t)$  into the point $x=\ell$ at $t=1$. The collapse is driven by the \emph{negative} pressure $P\left(\rho\right)=-\rho^{2}/2$ of this effective gas \cite{MKV}, and the solution has compact support \cite{KK2009,MKV}.  Once this hydrodynamic problem is solved, $h(x,t)$ can be found from the relation
\begin{equation}
h\left(x,t\right)=\int_{0}^{x}V\left(x',t\right)dx'+\int_{0}^{t}\rho\left(x=0,t'\right)dt',
\label{eq: h from V}
\end{equation}
where we have used Eq. \eqref{eq: OFM eq for h}, with the diffusion
term neglected, at $x=0$, and Eq.~\eqref{eq: reflecting condition}.
The inviscid hydrodynamic problem has an additional scale invariance property  \cite{MKV} which reduces the number of the dimensionless parameters to one.  Indeed, the rescaling tranformation
\begin{align}
&x'=\frac{x}{\Lambda^{1/3}}, \quad t'=t,  \quad 
\rho'\left(x',t\right)=\frac{\rho\left(x,t\right)}{\Lambda^{2/3}}, \nonumber \\
& V'\left(x',t'\right)=\frac{V\left(x,t\right)}{\Lambda^{1/3}}, \quad h'\left(x',t'\right)=\frac{h\left(x,t\right)}{\Lambda^{2/3}},\label{eq: hydrodynamic rescaling}
\end{align}
keeps Eqs.~\eqref{eq: continutity equation}, \eqref{eq: momentum equation} and \eqref{eq: h from V}
and the homogeneous boundary conditions invariant. The final-time condition
\eqref{eq: rho final time condition},
becomes
\begin{equation}
\rho'\left(x',t'=1\right)=\delta\left(x'-\ell'\right),\label{eq: rescaled final time condition}
\end{equation}
where $\ell'=\ell/\Lambda^{1/3}$ is the only parameter remaining in the problem. Alternatively, we can choose
$\ell/\sqrt{H}$ as the single parameter. One way of showing it is the following. Performing the rescalings \eqref{eq: hydrodynamic rescaling} in Eq. ~\eqref{eq: s in terms of rho}, we obtain
\begin{equation}\label{eq: action rescaling}
s\left(H,\ell\right)=  \Lambda^{5/3}s'\left(\ell'\right).
\end{equation}
Using this equation and the last relation in Eqs. \eqref{eq: hydrodynamic rescaling}, we obtain
\begin{align}
\frac{s\left(H,\ell\right)}{H^{5/2}} & =\frac{s'\left(\ell'\right)}{H'\left(\ell'\right)^{5/2}},\label{eq: rescaled ratio S/H^(5/2)}\\
\frac{\ell}{\sqrt{H}} & =\frac{\ell'}{\sqrt{H'\left(\ell'\right)}},\label{eq: rescaled ration l/sqrt(H)}
\end{align}
where $s'(\ell')$ is the rescaled action and $H'\left(\ell'\right)=h'\left(x'=\ell',t'=1\right)$ is the rescaled height.
Eqs.~\eqref{eq: rescaled ratio S/H^(5/2)} and \eqref{eq: rescaled ration l/sqrt(H)} yield
\begin{equation}
s\left(H\to\infty,\ell\right)=H^{5/2}f_{+}\left(\frac{\ell}{\sqrt{H}}\right),\label{eq: action scaling H->infty}
\end{equation}
where $f_{+}$ is to be found \cite{sameresult}.
Until the end of this section we will use the rescaled variables and omit the primes.

The solution to the rescaled \emph{full-line} problem involves a gas
cloud with an initial size of $4\sqrt{2H}/\pi$ which collapses symmetrically
into its center as $t\to1$ \cite{MKV}. Let us consider the main properties of the optimal path, as described by the inviscid Eqs.~(\ref{eq: continutity equation}) and  (\ref{eq: momentum equation}). If, in the half-line problem,
$\ell$ is larger than half this initial size,  $2\sqrt{2H}/\pi$, the same gas cloud, centered at  $x=\ell$, fits into
the interval $\left[0,2\ell\right]$, and the solution is just a full-line solution shifted  in space.
For  $\ell$ smaller than $2\sqrt{2H}/\pi$, the character of the solution changes, and we should expect a dynamical
phase transition at
\begin{equation}
\ell_{\text{cr}}\left(H\right)=\frac{2\sqrt{2}}{\pi}\sqrt{H}.\label{eq: l_cr(H)}
\end{equation}
Furthermore, for $\ell<\ell_{\text{cr}}$, the gas cloud must detach from the
reflecting boundary at $x=0$ at a finite time $0<t_{\star}<1$,
before collapsing into the point $x=\ell$ at $t=1$. The detachment time $t_{\star}$ is uniquely
determined by $\ell$: the larger $\ell$ is at fixed $H$, the closer
$t_{\star}$ will be to zero.

As in the full-line problem, the gas cloud  here has compact support at all
times: $x_{\text{l}}\left(t\right)<x<x_{\text{r}}\left(t\right)$, where $x_{\text{l}}$ and $x_{\text{r}}$ are the edges of support. For $\ell>\ell_{\text{cr}}\left(H\right)$,
$x_{\text{l}}\left(t\right)>0$ at all times, while for $\ell<\ell_{\text{cr}}\left(H\right)$
$x_{\text{l}}\left(t\right)>0$ at $t>t_{\star}$, and $x_{\text{l}}\left(t\right)=0$ for $0<t<t_{\star}$.

The gas density $\rho(x,t)$ and velocity $V(x,t)$ vanish
for $x>x_{\text{r}}\left(t=0\right)= x_{\text{r}}^{0}$ and $x<x_{\text{l}}\left(t=0\right)= x_{\text{l}}^{0}$
at all times.
The density vanishes identically on the intervals  $x_{\text{r}}\left(t\right)<x<x_{\text{r}}^{0}$ and $x_{\text{l}}^{0}<x<x_{\text{l}}\left(t\right)$, and the dynamics of
the velocity there is described by the Hopf equation (\ref{eq: Hopf equation}).
We call these regions the Hopf regions, and the region
$x_{\text{l}}\left(t\right)<x<x_{\text{r}}\left(t\right)$
the pressure-driven flow region, or simply the pressure flow region. Fig. \ref{fig: flow regions} below shows
the boundaries of these regions in the $\left(x,t\right)$ plane.

Here is a plan for the remainder of this section. By transforming from the Eulerian coordinate $x$ to the
Lagrangian mass coordinate, we will reduce the set of equations \eqref{eq: continutity equation} and \eqref{eq: momentum equation} to a single nonlinear elliptic equation of the second order, and solve it
numerically. We will indeed find the two different regimes of the most probable paths
and the large deviation function, depending on the parameter $\ell/\sqrt{H}$, and the ensuing phase transition.

\subsection{Lagrangian coordinates and numerical method}
\label{subsec:Lagrangian-coordinates}

To our knowledge, at $\ell<\ell_{\text{cr}}\left(H\right)$,
the inviscid hydrodynamic problem  cannot be solved
analytically, and we resort to numerical calculations. A numerical scheme which uses
the Eulerian $x$ coordinate
cannot be efficient, as an increasingly finer resolution near
the location of the collapse $x=\ell$ would be needed in order to resolve the dynamics
with sufficient precision. Using a Lagrangian coordinate is more suitable, as
small features, which develop along the $x$ coordinate,
are spread more evenly along a Lagrangian coordinate.

Since the total mass is conserved, see Eq.~\eqref{eq: rho conservation law},
it is convenient to use the Lagrangian mass coordinate \cite{ZR}, defined by
\begin{equation}
m\left(x,t\right)=\int_{0}^{x}\rho\left(x',t\right)dx'.\label{eq: mass coordinate definition}
\end{equation}
The inverse relation is
\begin{equation}
x\left(m,t\right)=\int_{0}^{m}\frac{dm'}{\rho\left(m',t\right)}+\int_{0}^{t}V\left(m=0,t'\right)dt',\label{eq: x(m,t)}
\end{equation}
where we used the fact that the Lagrangian time derivative relates the velocity and position of a gas parcel by
\begin{equation}
V\left(m,t\right)=\partial_{t}x\left(m,t\right), \label{eq: V and x relation}
\end{equation}
and the initial condition \eqref{eq: V initial condition}.

\begin{figure} [t]
\includegraphics[scale=0.5]{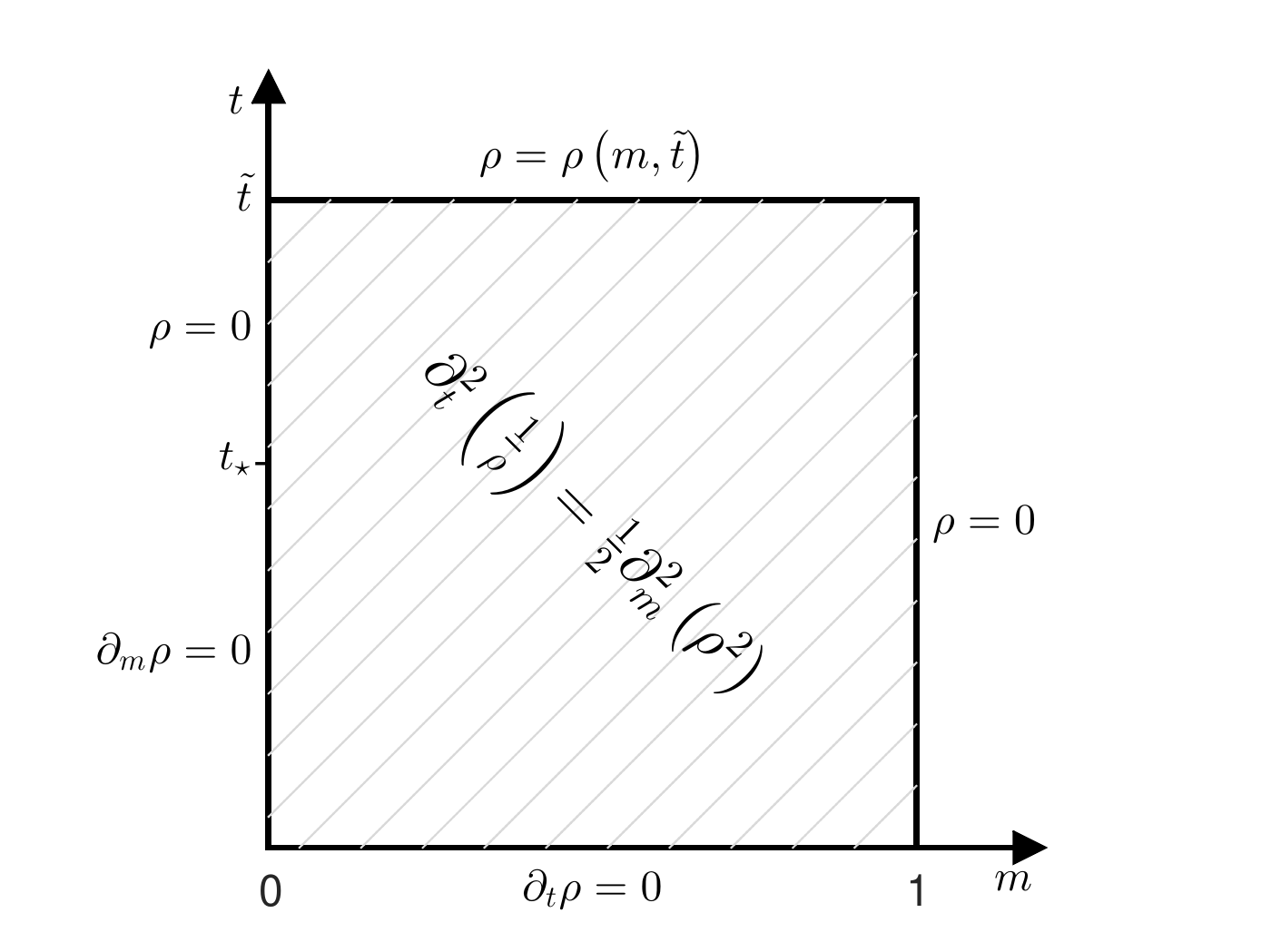}
\caption{\label{fig: single equation problem} The geometry and boundary conditions for
the pressure-driven flow in the Lagrangian mass coordinate.}
\end{figure}

In the Lagrangian representation, Eqs.~\eqref{eq: continutity equation}
and \eqref{eq: momentum equation} in the pressure region take the form
\begin{align}
\partial_{m}V & =\partial_{t}\left(\frac{1}{\rho}\right),\label{eq: Lagrangian continutity eq}\\
\partial_{t}V & =\frac{1}{2}\partial_{m}\left(\rho^{2}\right).\label{eq: Lagrangian momentum eq}
\end{align}
By differentiating Eq. \eqref{eq: Lagrangian continutity eq} with
respect to $t$ and Eq. \eqref{eq: Lagrangian momentum eq} with respect
to $m$, we eliminate $V$ and arrive at a single nonlinear partial differential equation for $\rho(m,t)$:
\begin{equation}
\partial_{t}^{2}\left(\frac{1}{\rho}\right)
=\frac{1}{2}\partial_{m}^{2}\left(\rho^{2}\right).\label{eq: single equation for rho}
\end{equation}
As the total mass of the gas is conserved and equal to $1$  [see Eq.~(\ref{eq: rescaled final time condition})], Eq.~(\ref{eq: single equation for rho})  should be solved inside the square $(0,1)\times (0,1)$
of the $(m,t)$ plane, see Fig.~\ref{fig: single equation problem}.

What are the boundary conditions for the elliptic equation (\ref{eq: single equation for rho})? Using Eq.~\eqref{eq: Lagrangian continutity eq}, we transform the initial condition \eqref{eq: V initial condition}
to
\begin{equation}
\partial_{t}\rho\left(m,t=0\right)=0.\label{eq: t=00003D0 BC}
\end{equation}
The boundary condition at $m=1$ is
\begin{equation}
\rho\left(m=1,t\right)=0.\label{eq: m=00003D1 BC}
\end{equation}
The boundary condition at $m=0$ is a bit more involved. The parameter $\ell$ affects the problem only
via the detachment time $t_{\star}$ (see the paragraph after Eq.~\eqref{eq: t=00003Dt_tilde BC} below). Therefore,  it is convenient to reparameterize the problem in terms of $t_{\star}$ instead of $\ell$.  For $t\le t_{\star}$ the gas density
at $x=m=0$ is nonzero. Then, using the relation $dm=\rho\left(x,t\right)dx$,
we can transform the reflecting condition \eqref{eq: reflecting condition}
to $\partial_{m}\rho\left(m=0\right)=0$. For $t>t_{\star}$ the gas density is zero at
$m=0$. Overall, the boundary condition  at $m=0$ is
\begin{equation}
\begin{cases}
\partial_{m}\rho\left(m=0,t\right)=0, & t\le t_{\star},\\
\rho\left(m=0,t\right)=0, & t>t_{\star}.
\end{cases}\label{eq: m=00003D0 BC}
\end{equation}

The last boundary condition follows from the final-time condition~\eqref{eq: rho final time condition}.
As the latter involves a delta-function, the Lagrangian mass coordinate is degenerate at $t=1$.
We overcame this difficulty by exploiting the fact that, very close to $t=1$, the hydrodynamic solution (1) behaves as the full-line solution centered at $x=\ell$, and (2) exhibits self-similarity. Using the results of Ref. \cite{MKV}, this self-similar asymptotic can be written as
\begin{equation}
\frac{\rho_{\text{ss}}\left(x,t\right)}{r(t)} = \begin{cases}
1-\frac{16}{9}r^{2}\left(t\right)\left(x-\ell\right)^{2}, & |x-\ell|\le \frac{3}{4r(t)},\\
0, & |x-\ell|\ge \frac{3}{4r(t)},
\end{cases}\label{eq: full line rho}
\end{equation}
where
\begin{equation}
r(t)=\left[4\left(1-t\right)\right]^{-2/3}.\label{eq: r(t=00005Cto1)}
\end{equation}
Therefore, we can solve the problem numerically only until a time $\tilde{t}$ sufficiently close to $1$, and use the similarity solution for
$\tilde{t}\le t\le1$. In the numerical solution we enforce a final-time condition
at $t=\tilde{t}$ by setting the gas density $\rho_{\text{ss}}(x,\tilde{t})$ from Eqs.~(\ref{eq: full line rho}) and (\ref{eq: r(t=00005Cto1)}). What is left is to transform  $\rho_{\text{ss}}(x,\tilde{t})$ to the Lagrangian mass coordinate. Let us denote for brevity $\tilde{r}=r\left(\tilde{t}\right)$.  According
to Eq. \eqref{eq: mass coordinate definition}, the mass coordinate
at $t=\tilde{t}$ is
\begin{equation}
m\left(x,\tilde{t}\right)=\frac{1}{2}
+\tilde{r}\left(x-\ell\right)-\frac{16\tilde{r}^{3}}{27}\left(x-\ell\right)^{3}.\label{eq: m(x,t_tilde)}
\end{equation}
Inverting this relation requires solving a cubic equation, which is conveniently done in a parametric form:
\begin{equation}
x\left(m,\tilde{t}\right)
=\ell
+\frac{3}{4\tilde{r}}\left[\cos \frac{\theta\left(m\right)}{3}
-\sqrt{3}\sin \frac{\theta\left(m\right)}{3}\right]\label{eq: x(m,t_tilde)}
\end{equation}
where
\begin{equation}
\theta\left(m\right)=\arctan\left(2m-1,2\sqrt{m-m^{2}}\right),\label{eq: theta(m)}
\end{equation}
and the function $\arctan(x,y)$ gives the arc tangent of $y/x$, taking into account which quadrant the point $(x,y)$ is in \cite{Wolfram}.
Plugging Eq.~(\ref{eq: x(m,t_tilde)}) back in Eq. \eqref{eq: full line rho}, we arrive at
the final-time condition in the Lagrangian representation
\begin{equation}
\rho\left(m,\tilde{t}\right)=\tilde{r}\left[\sqrt{3}\sin \frac{2\theta\left(m\right)}{3}
-2\sin^{2} \frac{\theta\left(m\right)}{3}\right] .
\label{eq: t=00003Dt_tilde BC}
\end{equation}
\begin{figure} [t]
\includegraphics[scale=0.5]{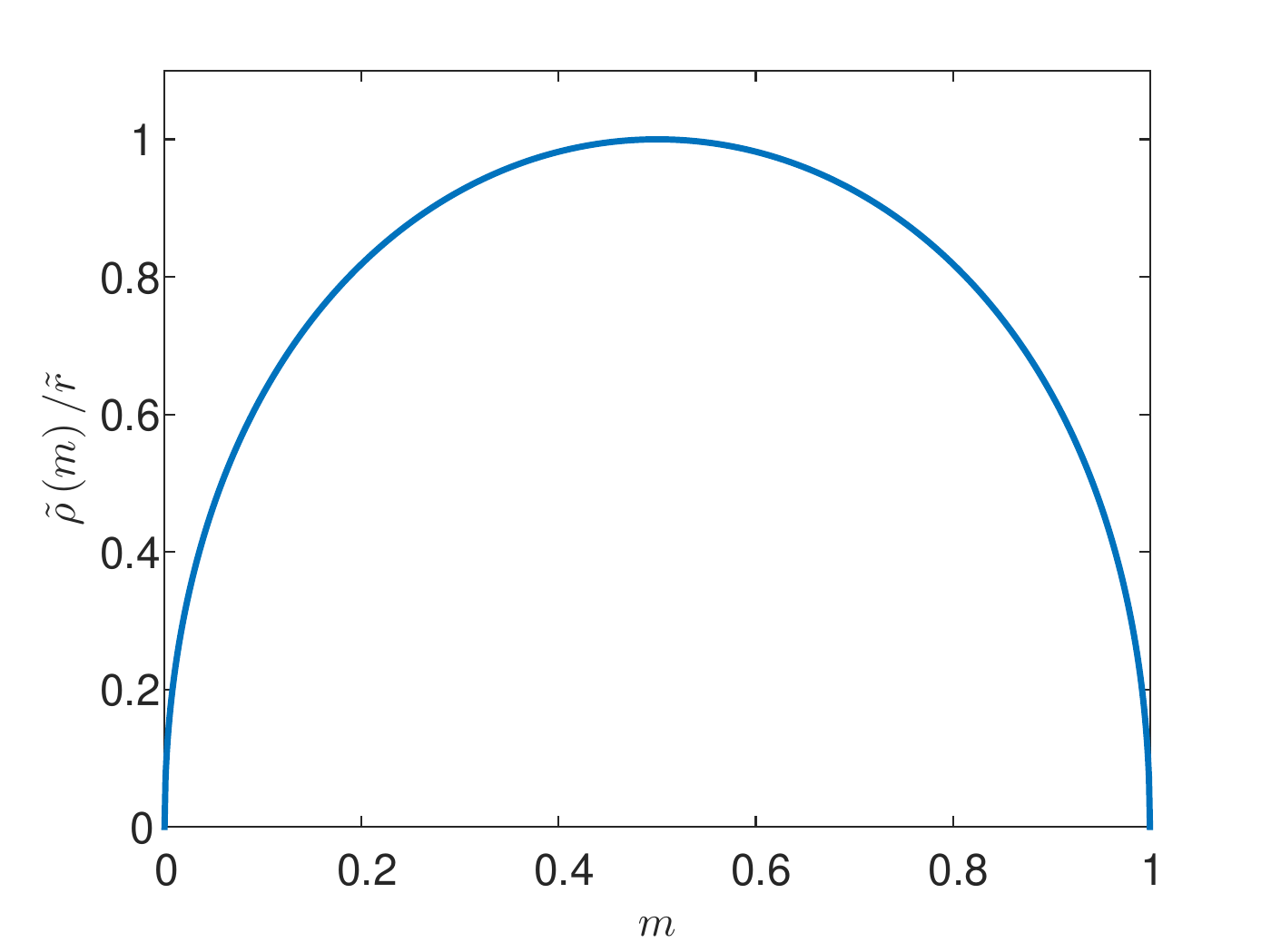}
\caption{The final-time condition (\ref{eq: t=00003Dt_tilde BC}) for the pressure flow in the Lagrangian coordinate.}
\label{fig:lagrangian_final_time}
\end{figure}
The function $\rho\left(m,\tilde{t}\right)$ is shown in Fig. \ref{fig:lagrangian_final_time}.
As one can see, a very narrow density profile in the Eulerian coordinate (which would be a delta-function at $\tilde{t}=1$)  gives way to a broad function in the Lagrangian coordinate. This is clearly advantageous for numerical calculations.
Importantly, Eq. (\ref{eq: t=00003Dt_tilde BC}) does not depend on $\ell$. It is precisely this fact that enables us to reparameterize the problem in terms of  the detachment time $t_{\star}$. Using the reparametrization,  we compute the Eulerian collapse location $x=\ell$ at $t=\tilde{t}$ for each specified value of $t_{\star}$.
The geometry and boundary conditions for the pressure-driven flow in the Lagrangian representation are
shown in Fig. \ref{fig: single equation problem}.

We use Newton's method \cite{Mazumder}
to solve Eq.~(\ref{eq: single equation for rho}) for $\rho\left(m,t\right)$, typically
with $\tilde{r}=750$, which corresponds to $1-\tilde{t}\simeq 1.2 \cdot10^{-5}$. The rapid growth of $\rho\left(m,t\right)$ as $t$ approaches $\tilde{t}$, see Eq. ~(\ref{eq: r(t=00005Cto1)}),  causes a numerical difficulty. We overcame it by using a non-uniform mesh, see Appendix \ref{Appendix: numerical scheme H=00005Cto=00005Cinfty}. Then,
using the numerical solution of Eq.~\eqref{eq: Lagrangian momentum eq}, we find $V(m,t)$. With $\rho\left(m,t\right)$ and $V\left(m,t\right)$ at hand, we transform the pressure flow solution back to the Eulerian coordinate using
Eq. \eqref{eq: x(m,t)}.

\begin{figure}
\centering{}\includegraphics[scale=0.5]{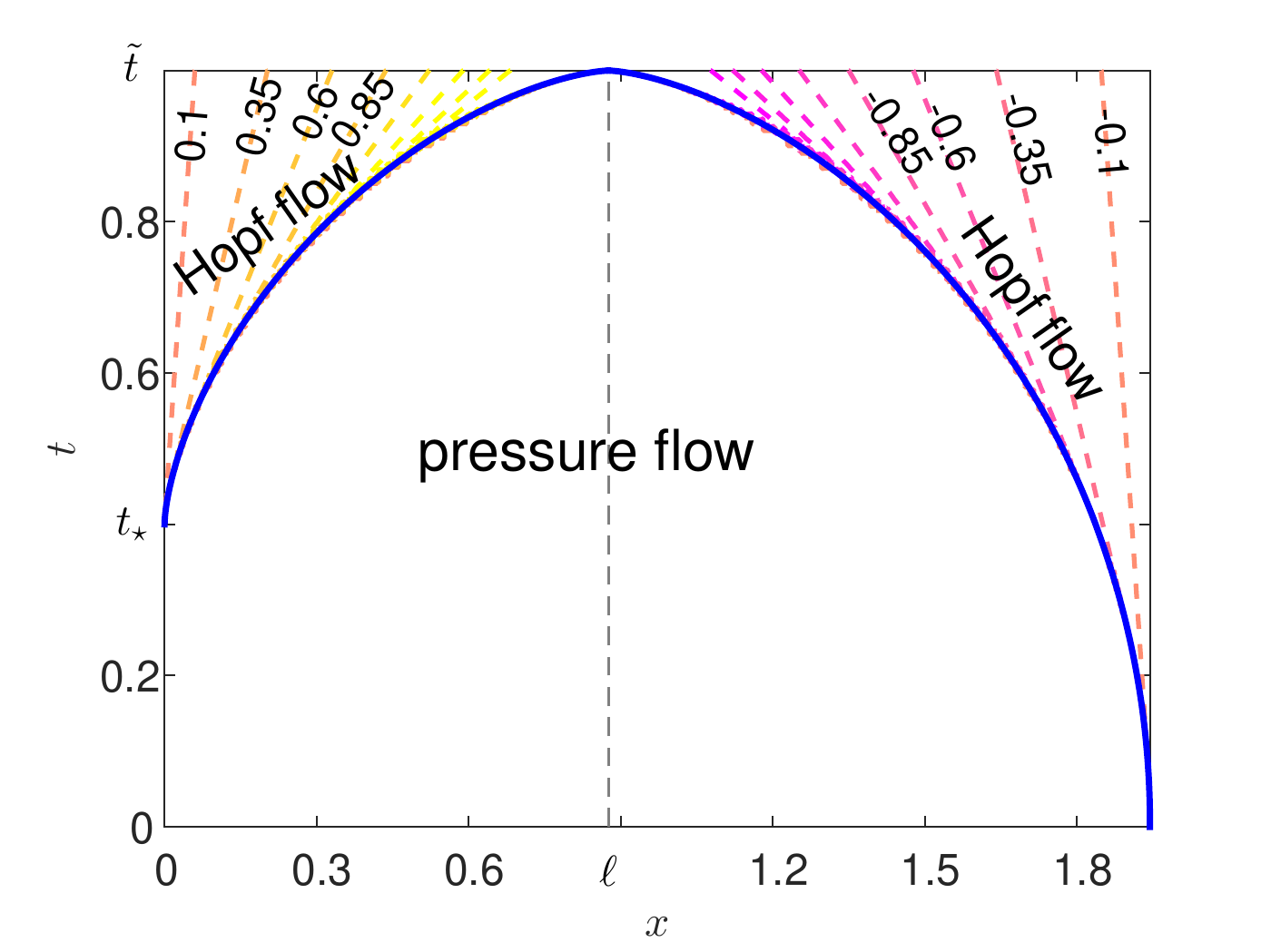}
\caption{The flow regions of the effective hydrodynamic problem in the Eulerian coordinate for $\ell\simeq 0.876$
(or $t_{\star}=0.4$). The solid lines to the left and right of $x=\ell$
are the edges of the compact support of the pressure flow region, $x_{\text{l}}\left(t\right)$
and $x_{\text{r}}\left(t\right)$, respectively.  $x_{\text{r}}\left(t\right)$
decreases as a function of time for any $t>0$, while $x_{\text{l}}\left(t\right)$ increases only
for $t>t_{\star}$, after the gas cloud detaches from $x=0$. The pressure
flow region shrinks to zero at $t=1$, as the gas collapses to $x=\ell$.
The dashed lines are characteristics of the Hopf equation \eqref{eq: Hopf equation},
emanating from the edges of the pressure flow region and carrying with
them constant values of the velocity $V(x,t)$ into the Hopf regions. Some of these constant  values are indicated.}
\label{fig: flow regions}
\end{figure}

To compute $V(x,t)$ in the regions of Hopf flow, see Eq. (\ref{eq: Hopf equation}), we implemented numerically the matching procedure
of Ref.  \cite{MKV}. Using numerical characteristics, we match the implicit general solution to the Hopf equation
\cite{LandauLifshitzFluidMechanics},
\begin{equation}
V=F\left(x-Vt\right),\label{eq: Hopf eq general solution}
\end{equation}
with $V$ at the edges of the pressure flow region $x_{\text{l}}\left(t\right)$
and $x_{\text{r}}\left(t\right)$, see Fig. \ref{fig: flow regions}. Lastly,
we numerically evaluate the integrals over $x$ and $t$ in Eq. \eqref{eq: h from V} to find $h(x,t)$.
The choice of
mesh in $m$ and $t$ in the pressure flow region, and a brief description of the method of numerical characteristics
in the Hopf regions, are presented in Appendix \ref{Appendix: numerical scheme H=00005Cto=00005Cinfty}.

The final step is to compute $s$ and $H=h\left(\ell,1\right)$,
using Eqs.~\eqref{eq: s in terms of rho} and \eqref{eq: h from V} at $x=\ell$, respectively. We split the
integrals over time  into two regions,
$t\in\left[0,\tilde{t}\right]$ and $t\in\left[\tilde{t},1\right]$, and use the numerical solution in the former region, and the self-similar asymptotic (\ref{eq: full line rho}) in the latter one.

\subsection{Numerical results}
\label{subsec:Numerical-results}

We tested our numerical method by comparing its results at the critical point $t_{\star}=0$, when the boundary
at $x=0$ still has no effect, with analytical full-line results
\cite{MKV}:
\begin{equation}\label{analyticHD}
H=\frac{1}{2}\left(\frac{3\pi}{2}\right)^{2/3},\; s=\frac{1}{5}\left(\frac{3\pi}{2}\right)^{2/3},\;  \ell = \left(\frac{2\sqrt{3}}{\pi}\right)^{2/3}
\end{equation}
(in the rescaled units where $\Lambda=1$).  In this case $\ell$ is half the initial width of the gas cloud.  We found that the numerical and analytical results for $s$, $H$ and $\ell$ agree within less than $0.5\%$. Decreasing the mesh spacing by a factor of $1.5$ and $2$ for the $m$- and $t$-mesh, respectively, changed these results only by about $0.1\%$.

For $t_{\star}\ll1$, the effect of the boundary condition at $x=0$
is small, and the numerical density and velocity profiles are close to
the full-line profiles (a parabolic profile for the density, and a straight-line profile for
the velocity in the pressure flow region \cite{MKV}).

\begin{figure}
\includegraphics[scale=0.45]{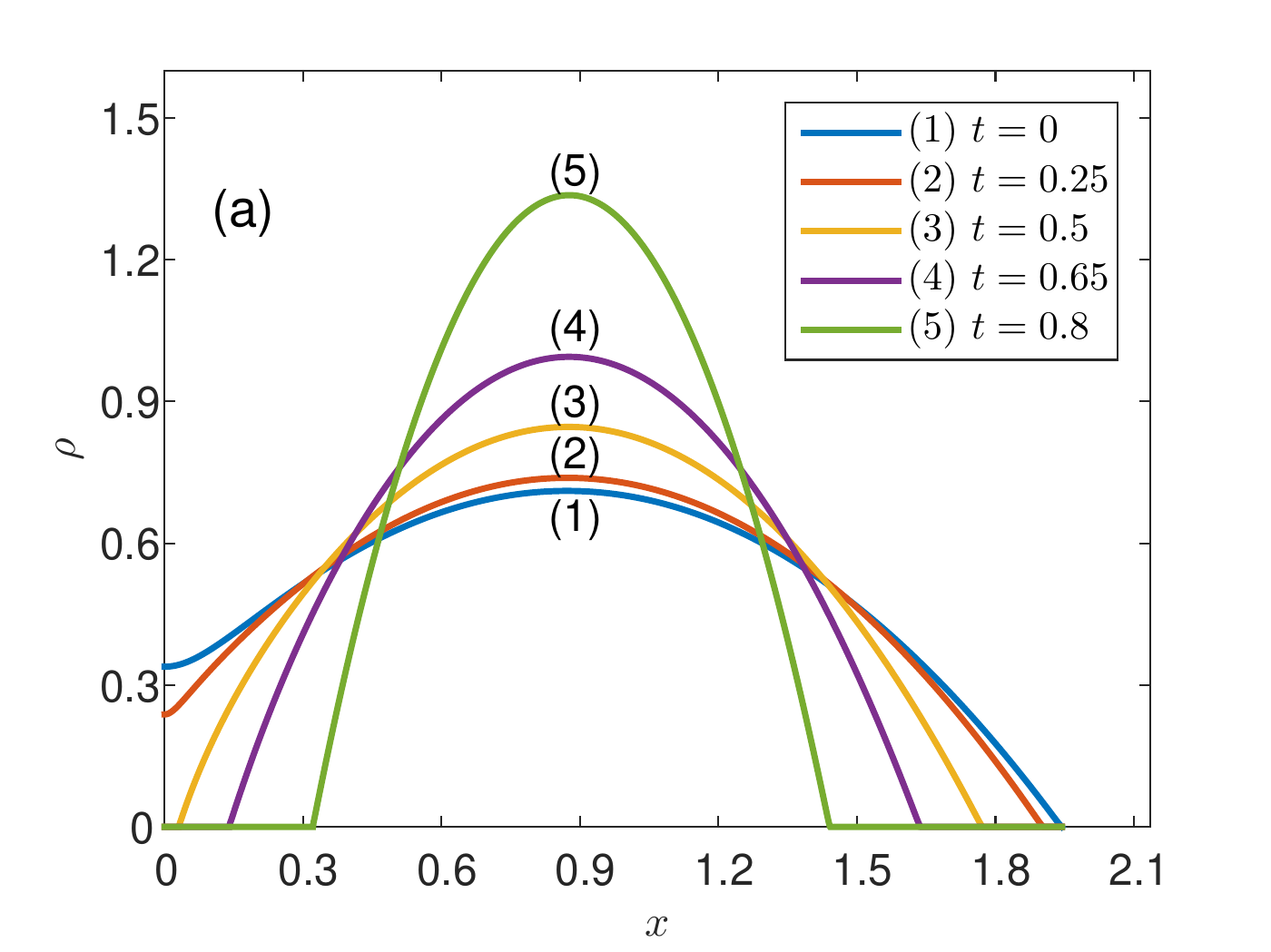}
\includegraphics[scale=0.45]{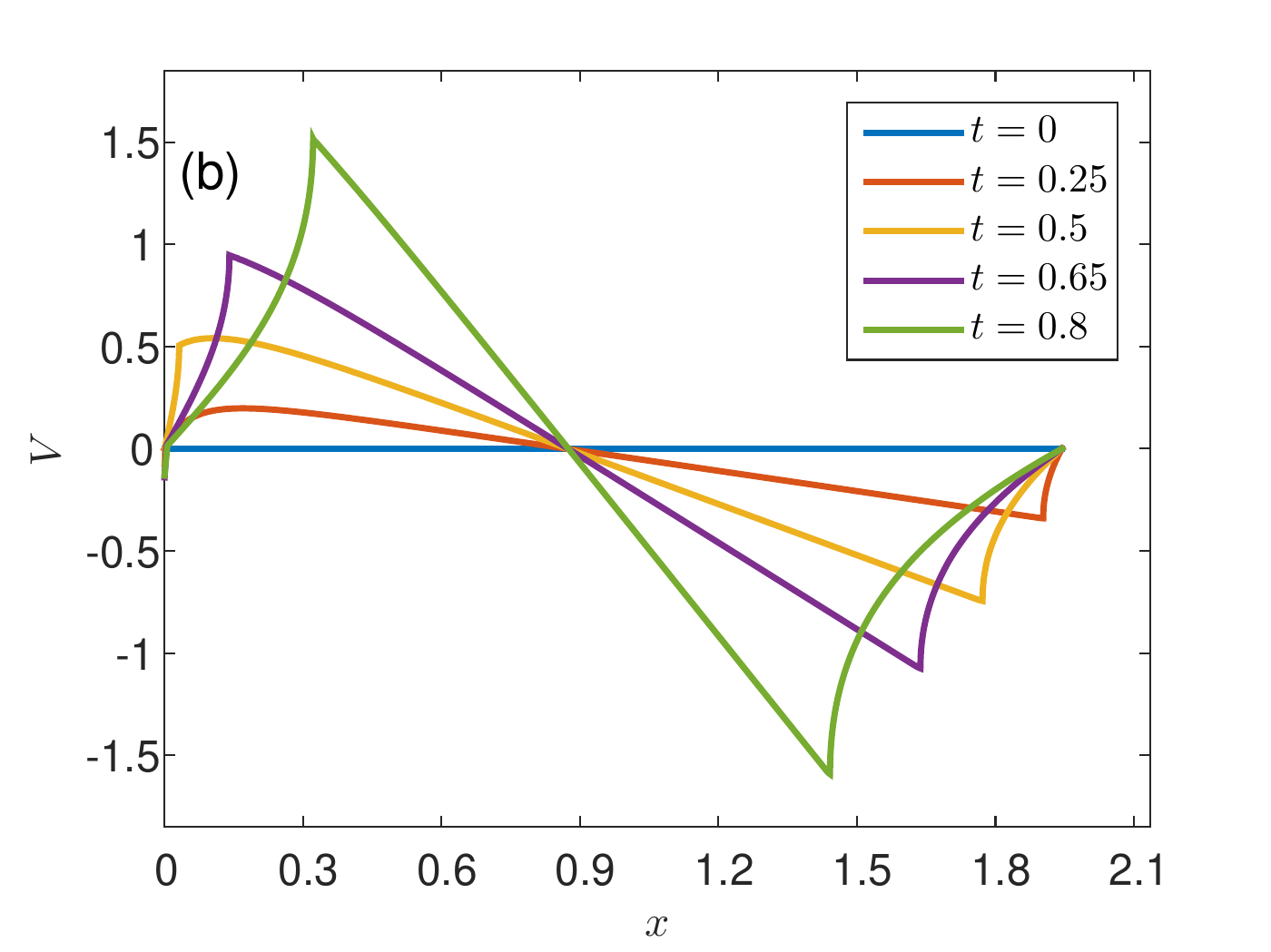}
\includegraphics[scale=0.45]{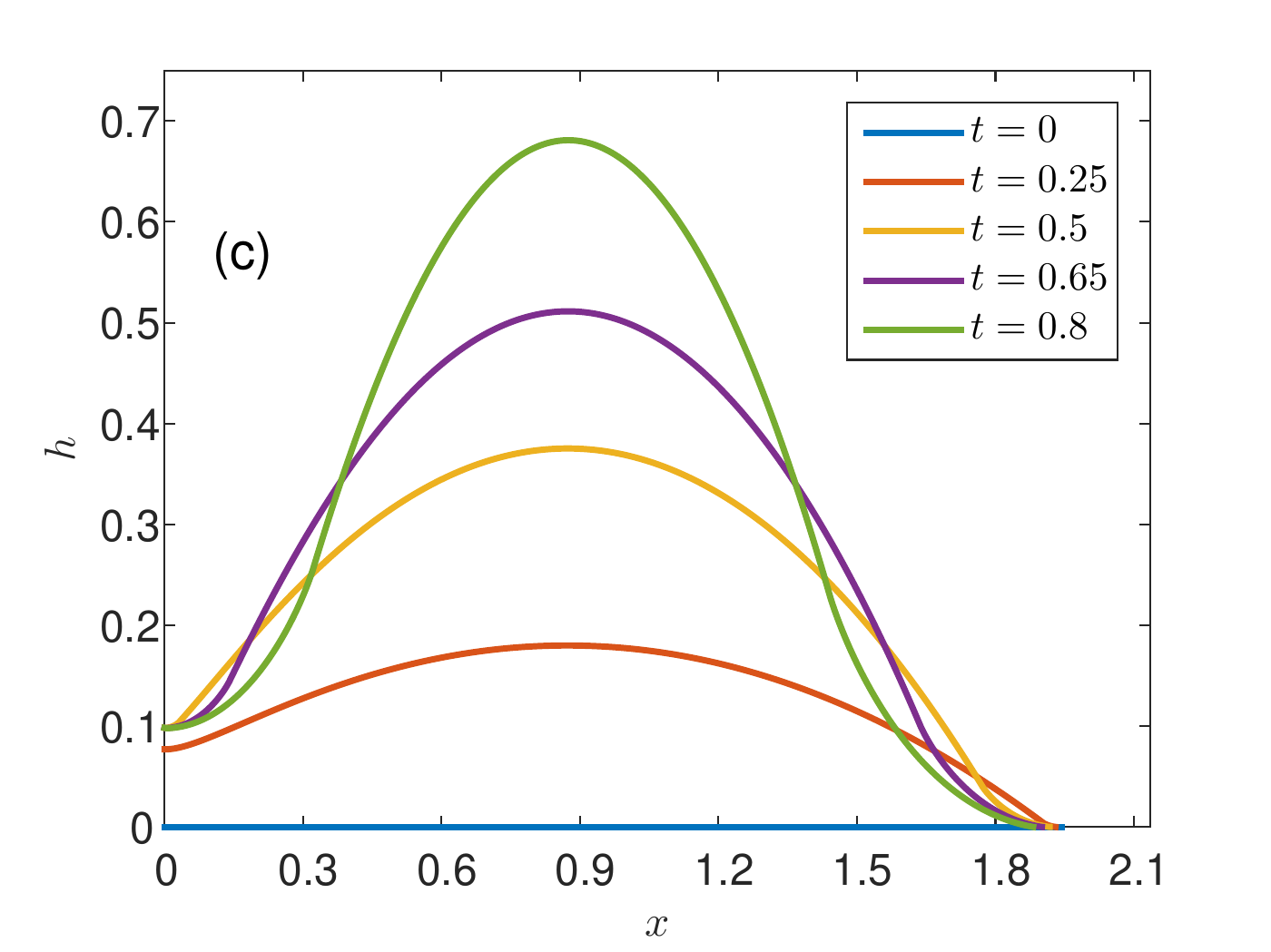}
\caption{The numerically found optimal path of the system, corresponding to the $H\gg1$ tail for $t_{\star}=0.4$ (or $\ell=0.876$). Shown are the spatial profiles of $\rho$ (a), $V$ (b) and $h$ (c) at times
$t=0, 0.25, 0.5, 0.65$ and $0.8$.  At $t=t_{\star}$ the gas detaches from  $x=0$. The regions of pressure flow and Hopf flow are clearly seen in panel (b). As $t$ approaches $\tilde{t}$, the pressure flow solution
converges to the (self-similar asymptotic of) the full-line solution of Ref. \cite{MKV}. To remind the reader, $x, t, \rho, V$ and $h$ scale with $\Lambda$ as stated in Eq.~(\ref{eq: hydrodynamic rescaling}).}
\label{fig: H>>1 numeric t*=00003D0.4 profiles}
\end{figure}

Larger values of $t_{\star}$ (that is, smaller values of $\ell$) lead to more complicated dynamics, see Fig. \ref{fig: H>>1 numeric t*=00003D0.4 profiles}.
Still, well after the gas detaches from $x=0$, the
numerical solution approaches the $t\to 1$ asymptotic of the full-line solution, in agreement with our initial expectations.

\begin{figure}
\centering{}\includegraphics[scale=0.5]{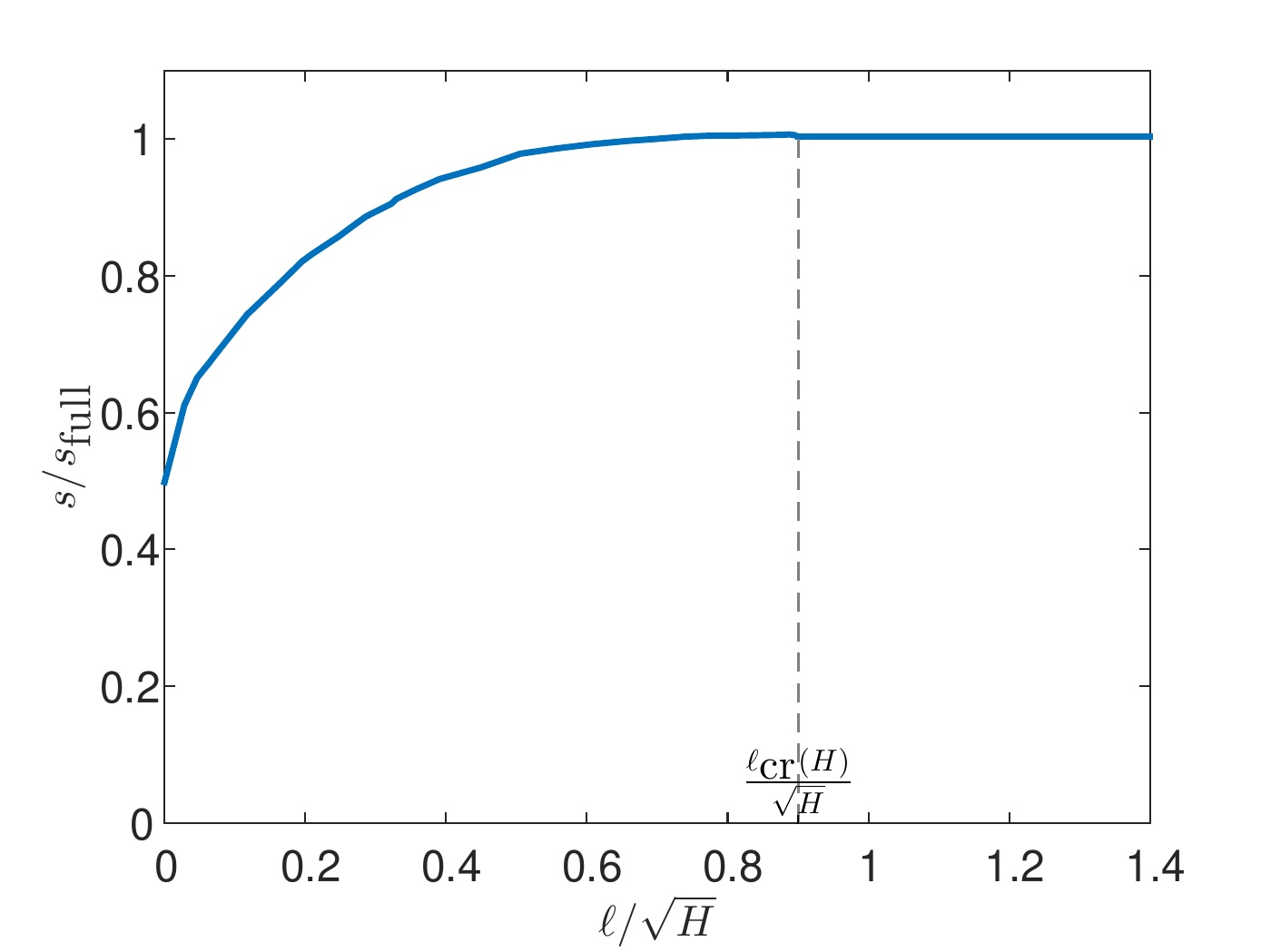}\caption{Numerical results for $s/s_{\text{full}}$ in the $H\gg1$ tail as
a function of $\ell/\sqrt{H}$. Up to the factor $8\sqrt{2}/15\pi$, $s/s_{\text{full}}$
is the function $f_{+}\left(\ell/\sqrt{H}\right)$, see Eqs. \eqref{eq: action scaling H->infty}
and \eqref{eq: H>>1 full line action}. Evident is a gradulal crossover from $s=s_{\text{full}}/2$
at $\ell=0$ to $s=s_{\text{full}}$ at $\ell=\ell_{\text{cr}}\left(H\right)$, and a sharp transition at $\ell=\ell_{\text{cr}}\left(H\right)$. For $\ell>\ell_{\text{cr}}\left(H\right)$ the action is
independent of $\ell$.  The numerical value of  $\ell_{\text{cr}}\left(H\right)/\sqrt{H}$ agrees with
$2\sqrt{2}/\pi$ up to less than $0.1\%$.}
\label{fig: H>>1 s(L/sqrt(H))}
\end{figure}

Fig. \ref{fig: H>>1 s(L/sqrt(H))} shows our numerical results for
the action, in the units of the full-line action \cite{KK2009,MKV}
\begin{equation}
s_{\text{full}}\left(H\gg 1\right)\simeq\frac{8\sqrt{2}}{15\pi}H^{5/2},\label{eq: H>>1 full line action}
\end{equation}
as a function of $\ell/\sqrt{H}$. The horizontal line at $\ell/\sqrt{H}>2\sqrt{2}/\pi$ [see Eq.~(\ref{eq: l_cr(H)})] is the numerical value for $t_\star=0$.
The numerical results satisfy the expected asymptotics
\begin{align}
s\left(H\gg1,\ell=0\right) & =\frac{1}{2}s_{\text{full}}\left(H\gg1\right),\label{eq: s(H>>1) =00005Cell=00003D0}\\
s\left(H\gg1,\ell\geq\ell_{\text{cr}}\left(H\right)\right) & = s_{\text{full}}\left(H\gg1\right),\label{eq: s(H>>1) large =00005Cell}
\end{align}
up to less than $0.5\%$ \cite{verified}. Also evident in Fig. \ref{fig: H>>1 s(L/sqrt(H))} is a phase transition at the same critical value $\ell/\sqrt{H}=2\sqrt{2}/\pi$ as  in the ring problem \cite{SMS2018}. Although the details of the hydrodynamic solution at $\ell/\sqrt{H}<2\sqrt{2}/\pi$ in these two problems are in general different, they are quite similar close to the transition. We believe, therefore, that the order of the phase transition in these two problems is the same: $5/2$. Unfortunately, the precision of our numerical solution in the vicinity of the phase transition is insufficient for a conclusive verification of this hypothesis, because of the high-order numerical derivatives of $s$ required in this calculation.

\section{Summary and discussion}
\label{Sec: summary}

The presence of an additional parameter $\ell=L/\sqrt{\nu t}$ leads to a rich phase diagram (see Fig. \ref{fig: phase diagram}) of scaling behaviors
of the height probability $\mathcal{P}\left(H,L,t\right)$ of the KPZ interface on the half-line. At small $|H|$, $\mathcal{P}\left(H,L,t\right)$ is a Gaussian, with a variance which
is $\ell$-dependent, see Fig. \ref{fig: linear action}. At large negative $H$, the distribution obeys the scaling behavior, described by Eq.~(\ref{eq: H to -=00005Cinfscaling}). The function $f_{-}$, which we calculated analytically, is shown in Fig. \ref{fig: negative tail action}. It describes a first-order dynamical phase transition, which results from a competition between two different histories of the system, conditioned on reaching the height $H$ at the point $x=L$.

At large positive $H$, the scaling behavior of $\mathcal{P}\left(H,L,t\right)$ is described by Eq.~(\ref{eq: H to =00005Cinfscaling}). The function $f_{+}$ is shown in Fig. \ref{fig: H>>1 s(L/sqrt(H))}. In order to compute it, we developed a numerical method which employs the Lagrangian mass coordinates and transforms the two coupled OFM equations into a single nonlinear second-order elliptic equation.    The function $f_{+}$  also describes a dynamical phase transition. Its mechanism, however, is different from that of the negative tail of the distribution. First, this transition is smoothed by small diffusion effects. Second, it appears when the effective ``gas cloud", describing the optimal history of the KPZ noise field, conditioned on $H$, starts ``feeling" the presence of the reflecting boundary at $x=0$. As this mechanism is very similar to the one in the ring problem \cite{SMS2018}, the order of the transition is apparently the same: $5/2$, but more analytical or numerical work is needed to test this hypothesis.

For sufficiently large $\ell =L/\sqrt{\nu T}$ (that is, in the right part of the phase diagram in Fig. \ref{fig: phase diagram}), each of the distribution tails has a double structure. The moderately far $H>0$ tail, $1\ll \left|\lambda\right|H/\nu \lesssim \ell^2$, coincides with the $H>0$ tail for the full line, whereas the very far $H>0$ tail, $\left|\lambda\right|H/\nu \gg \ell^2$, coincides with that for the half line. Similarly, the moderately far
$H<0$ tail, $1\ll \left|\lambda H\right|/\nu \lesssim \ell^2$, coincides with the $H<0$ tail for the full line, whereas the very far $H<0$ tail, $\left|\lambda H\right|/\nu \gg \ell^2$, coincides with that for the half line.

As in the previous works \citep{KK2007,KK2008,KK2009,MKV,KMSparabola,Janas2016,MSchmidt2017,MSV_3d, SMS2018,SKM2018,MV2018}, we made two approximations. The main approximation is the saddle-point evaluation of the KPZ path integral, leading to the OFM formulation. An additional approximation (different for each of the regimes of small, large positive, or large negative $H$) enabled us to
separately consider the typical fluctuations and the two tails.
It would be very interesting to find out whether the short-time distribution tails, that we have found in this work, persist (at sufficiently large $|H|$) at arbitrary times.

\section*{ACKNOWLEDGMENTS}

We are grateful to Naftali Smith for useful discussions. T.A. and
B.M. acknowledge financial support from the Israel Science Foundation
(grant No. 807/16).\bigskip{}
\bigskip{}

\appendix

\section{Evaluating the integral $I\left(x,t,x_{0}\right)$ in Eq.~\eqref{eq: integral definition}}

\label{Appendix: evaluating the integral}
Let us denote $a=4\left(t-s\right)$ and $b=4\left(1-s\right)$. The
integral becomes
\begin{equation}
I\left(x,t,x_{0}\right)=\frac{1}{\pi}\int_{0}^{t}\frac{ds}{\sqrt{ab}}\int_{-\infty}^{\infty}d\xi e^{-\left[\frac{\left(\xi-x\right)^{2}}{a}+\frac{\left(\xi-x_{0}\right)^{2}}{b}\right]}.\label{eq: integral in terms of a,b}
\end{equation}
The integral over $\xi$ is a Gaussian integral,
\begin{align}
\int_{-\infty}^{\infty}d\xi e^{-\left[\frac{\left(\xi-x\right)^{2}}{a}+\frac{\left(\xi-x_{0}\right)^{2}}{b}\right]} & =\sqrt{\frac{\pi ab}{a+b}}\exp\left[-\frac{\left(x-x_{0}\right)^{2}}{a+b}\right].\label{eq: =00005Cxiintegralsolved}
\end{align}
Plugging back the definitions of $a$ and $b$, we have
\begin{equation}
I\left(x,t,x_{0}\right)=\int_{0}^{t}ds\,
\frac{\exp\left[-\frac{\left(x-x_{0}\right)^{2}}
{4\left(1+t-2s\right)}\right]}{\sqrt{4\pi\left(1+t-2s\right)}}.
\label{eq: I after d=00005Cxiintegration}
\end{equation}
Introducing
\begin{equation}
\eta=\frac{x-x_{0}}{\sqrt{4\left(1+t-2s\right)}},\label{eq: =00005Cetadefinition}
\end{equation}
we bring the remaining integral to
\begin{equation}
I\left(x,t,x_{0}\right)=\frac{x-x_{0}}{4\sqrt{\pi}}\int_{\eta_{0}}^{\eta_{t}}d\eta\,\frac{e^{-\eta^{2}}}{\eta^{2}}\label{eq: I in terms of =00005Ceta}
\end{equation}
with $\eta_{0}=\left(x-x_{0}\right)/\sqrt{4\left(1+t\right)}$
and $\eta_{t}=\left(x-x_{0}\right)/\sqrt{4\left(1-t\right)}$.
Using the known integral
\[
\int dz\,\frac{e^{-z^{2}}}{z^{2}}=-\frac{e^{-z^{2}}}{z}-\sqrt{\pi}\erf\left(z\right)\equiv-f\left(z\right),
\]
we arrive at
Eq. \eqref{eq: integral solution} of the main text.

\section{Dynamic solution from exact multisoliton solutions}
\label{Appendix: dynamic from multisoliton}

\begin{figure} [h]
\centering{}\includegraphics[scale=0.45]{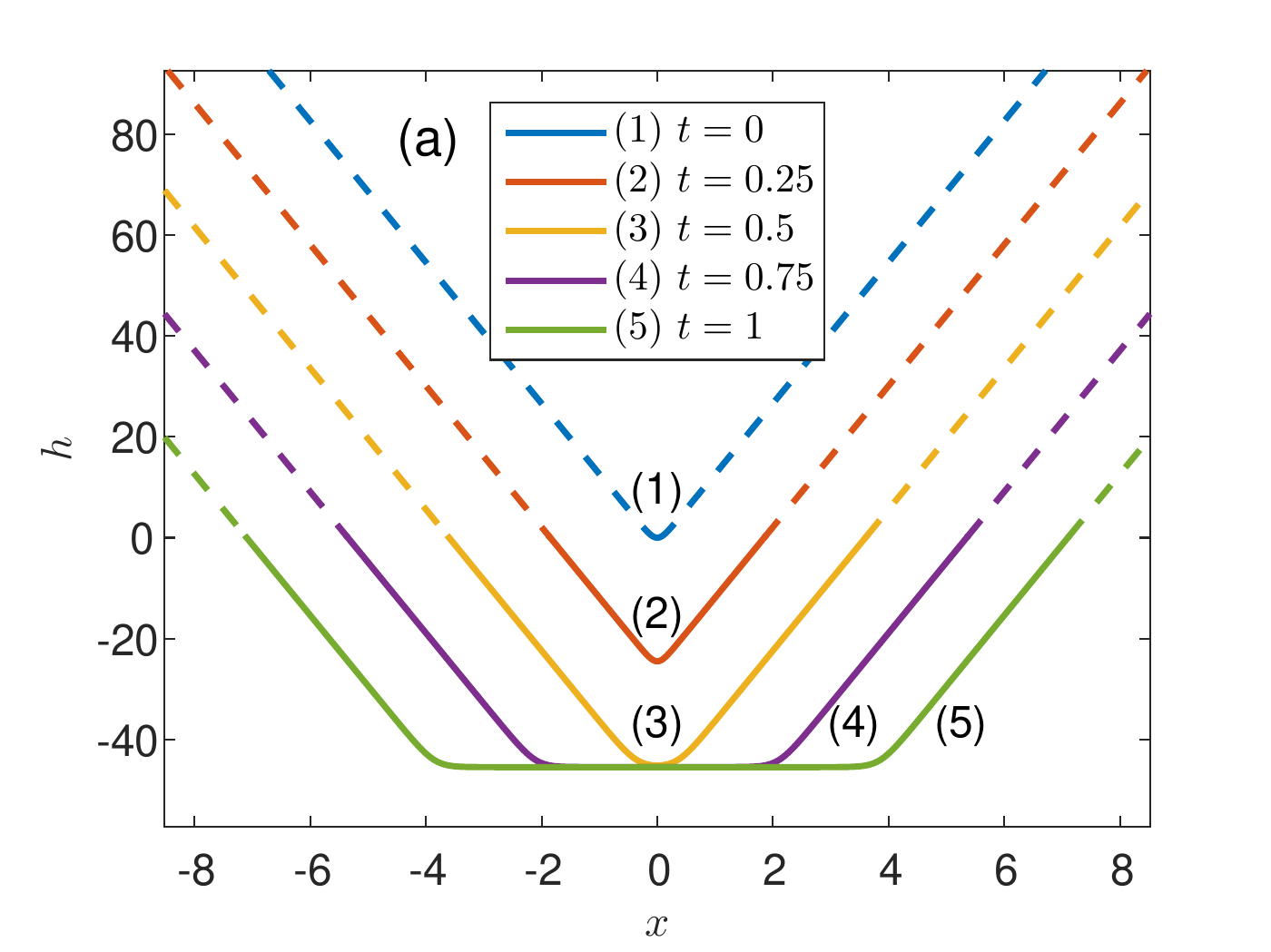}
\centering{}\includegraphics[scale=0.45]{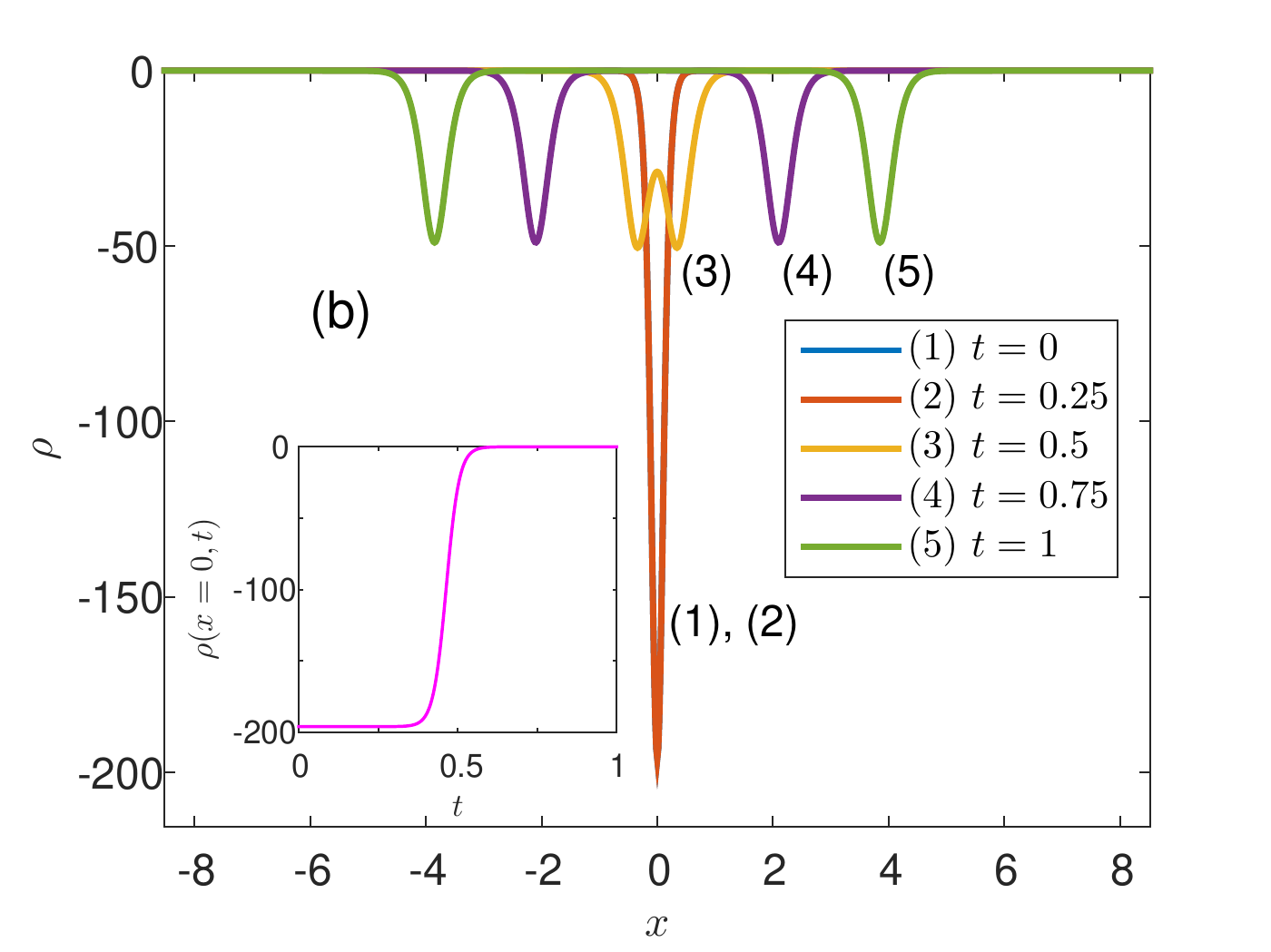}
\caption{An example of the two-soliton/two-ramp solution (\ref{eq: multisoliton h}) (a) and (\ref{eq: multisoliton rho}) (b) with $N=3$, $c_{1}=X_{1}=0$,
$c_{3}=-c_{2}$ and $X_{2}=-X_{3}=c_{2}\tau$, where $0<\tau<1$. The dashed lines in (a) indicate the non-physical parts of the solution that are replaced by the trivial solution $h=0$. The boundary layers, where the two solutions match, do not contribute, at leading order, to the action at large $|H|$.}
\label{fig:multisoliton}
\end{figure}

In Ref. \cite{Janas2016}
two families of multisoliton and multiramp solutions (for $\rho$ and
$h$, respectively) were found. The family relevant to this work is  given by
\begin{align}
\rho\left(x,t\right)= & -\frac{2\sum_{i,j=1}^{N}\left(c_{i}-c_{j}\right)^{2}e^{-c_{i}\left(c_{i}t+x-X_{i}\right)-c_{j}\left(c_{j}t+x-X_{j}\right)}}
{\left[\sum_{i=1}^{N}e^{-c_{i}\left(c_{i}t+x-X_{i}\right)}\right]^{2}},\label{eq: multisoliton rho}\\
h\left(x,t\right)= & -2\ln\left[\frac{4C}{\sum_{i=1}^{N}e^{-c_{i}\left(c_{i}t+x-X_{i}\right)}}\right].\label{eq: multisoliton h}
\end{align}
It holds for any integer $N>0$, and has $2N+1$ arbitrary constants:
$\left\{ c_{i},X_{i}\right\} _{i=1}^{N}$ and $C$.
The dynamic solution, described in Sec.~\ref{subsec: Dynamic-solution}, corresponds to $N=3$, $c_{1}=X_{1}=0$,
$c_{3}=-c_{2}$, and $X_{2}=-X_{3}=c_{2}\tau$. This solution is shown, for some choice of the parameters, in Fig. \ref{fig:multisoliton}.

In the limit of $-H\gg 1$
one has $c_1\gg 1$ and $c_2\gg 1$, and this multisoliton solution has two distinct asymptotics: the static soliton solution and two symmetric outgoing traveling soliton solutions, as shown in Fig.~\ref{fig:multisoliton} and described in
Sec.~\ref{subsec: Dynamic-solution}.

Going back to the two families of multisoliton and multiramp solution, discovered in Ref. \cite{Janas2016}, we note that
each of these families can be represented as a time-reversed version of the other.  This remarkable fact, previously unnoticed, is a consequence of a non-trivial time-reversal
symmetry of the OFM equations \eqref{eq: OFM eq for h} and \eqref{eq: OFM eq for rho} \cite{Canet2011, SM2018}.

\section{Numerical scheme for the $H\to\infty$ tail: more details}

\label{Appendix: numerical scheme H=00005Cto=00005Cinfty}

As $t$ approaches $\tilde{t}$, $\rho$
grows progressively fast, like $r(t)$, see Eq.~\eqref{eq: r(t=00005Cto1)}.
Therefore we chose an $r$-mesh with the number of points growing in a geometric
progression between $r_{0}$ and $\tilde{r}$ in $100$ steps. The
$t$ mesh is then found by setting a uniform mesh spacing of $\delta t=0.01$ for $t<0.7$, while for $t>0.7$ we compute $t\left(r\right)$ for every point
on the $r$-mesh, using Eq.~\eqref{eq: r(t=00005Cto1)}. The resulting $t$-mesh spacing decreases considerably
as $t$ grows. We restricted the maximum time step $\delta t$ to be no more than $0.01$ and used
a finer resolution of $\delta t=0.002$ around $t_{\star}$.

As for the $m$ mesh, we see from Eq. \eqref{eq: full line rho} that $\mu=4r\left(x-\ell\right)/3$
is a natural spatial coordinate for the density. Therefore, we used
a  mesh uniform in $\mu$ with $601$ divisions between $\mu=-1$ and
$\mu=1$. The $m$ mesh is computed from it by using  Eq. \eqref{eq: m(x,t_tilde)}:
\begin{equation}
m\left(\mu\right)=\frac{1}{2}+\frac{3}{4}\mu-\frac{1}{4}\mu^{3}.\label{eq: m(=00005Cmu)}
\end{equation}
The resulting $m$ mesh spacing, $\delta m$,  is small close to the edges of the pressure
flow region $m=0$ and $m=1$. As a function of $m$, $\delta m$ behaves as $\rho\left(m,\tilde{t}\right)$ shown in Fig. \ref{fig:lagrangian_final_time}, up to a scale factor of $0.025$. Our finite-difference approximation of the derivatives, used for the numerical solution of Eq. \eqref{eq: single equation for rho}, properly takes into account the non-uniformity of the mesh.

In the Hopf regions we use the fact
that the solution is constant along the characteristics
$x=Vt+\text{const}$ which are straight
lines.
Hence, once the velocity at the right edge
of the pressure flow region, $V\left(x=x_{\text{r}}\left(t\right),t\right)$,
is known, we can draw straight lines, with a slope $dx/dt=V_{j}= V\left(x_{\text{r}}\left(t_{j}\right),t_{j}\right)$, from each point $\left(x_{\text{r}}\left(t_{j}\right),t_{j}\right)$, and set the velocity along that line to be $V_{j} =\text{const}$. The same is done for the left edge $x_{\text{l}}\left(t\right)$. As a result, we have a
set of points in the $\left(x,t\right)$ plane with known
velocity, and determine the velocity at any other point in the Hopf region by linear interpolation.


\begin{thebibliography}{100}

\bibitem{KPZ} M. Kardar, G. Parisi, and Y.-C. Zhang, Phys. Rev. Lett.
\textbf{56}, 889 (1986).

\bibitem{signlambda} Changing the sign of $\lambda$ is equivalent
to changing $h$ to $-h$.

\bibitem{HHZ} T. Halpin-Healy and Y.-C. Zhang, Phys. Reports \textbf{254},
215 (1995); T. Halpin-Healy and K. A. Takeuchi, J. Stat. Phys. \textbf{160},
794 (2015).

\bibitem{Barabasi} A.-L. Barabasi and H. E. Stanley, \textit{Fractal
Concepts in Surface Growth} (Cambridge University Press, Cambridge,
UK, 1995).

\bibitem{Krug} J. Krug, Adv. Phys. \textbf{46}, 139 (1997).

\bibitem{Corwin} I. Corwin, Random Matrices: Theory Appl. \textbf{1},
1130001 (2012).

\bibitem{QS} J. Quastel and H. Spohn, J. Stat. Phys. \textbf{160},
965 (2015).

\bibitem{S2016} H. Spohn, in \textit{Stochastic Processes and Random
Matrices}, \textit{Lecture Notes of the Les Houches Summer School},
edited by G. Schehr, A. Altland, Y. V. Fyodorov and L. F. Cugliandolo
(Oxford University Press, Oxford, 2015), vol. 104.

\bibitem{Takeuchi2017} K. A. Takeuchi, Physica A \textbf{504}, 77
(2018).

\bibitem{displacement} One subtracts from the one-point surface height
the noise-induced systematic shift of the surface.

\bibitem{KK2007} I. V. Kolokolov and S. E. Korshunov, Phys. Rev.
B \textbf{75}, 140201(R) (2007).

\bibitem{KK2008} I. V. Kolokolov and S. E. Korshunov, Phys. Rev.
B \textbf{78}, 024206 (2008).

\bibitem{KK2009} I. V. Kolokolov and S. E. Korshunov, Phys. Rev.
B \textbf{80}, 031107 (2009).

\bibitem{MKV} B. Meerson, E. Katzav, and A. Vilenkin, Phys. Rev.
Lett. \textbf{116}, 070601 (2016).

\bibitem{Janas2016} M. Janas, A. Kamenev, and B. Meerson, Phys. Rev.
E \textbf{94}, 032133 (2016).

\bibitem{SKM2018} N. R. Smith, A. Kamenev and B. Meerson, Phys. Rev.
E \textbf{97}, 042130 (2018).


\bibitem{DMRS} P. Le Doussal, S. N. Majumdar, A. Rosso, and G. Schehr,
Phys. Rev. Lett. \textbf{117}, 070403 (2016).

\bibitem{LeDoussal2017} A. Krajenbrink and P. Le Doussal, Phys. Rev.
E \textbf{96}, 020102(R) (2017).

\bibitem{SM2018} N. R. Smith and B. Meerson, Phys. Rev. E \textbf{97},
052110 (2018).

\bibitem{SMP} P. V. Sasorov, B. Meerson, and S. Prolhac, J. Stat.
Mech. (2017) P063203.

\bibitem{MSchmidt2017} B. Meerson and J. Schmidt, J. Stat. Mech.
(2017) 103207.

\bibitem{Corwinetal2018} I. Corwin, P. Ghosal, A. Krajenbrink, P.
Le Doussal, and L.-C. Tsai, Phys. Rev. Lett. \textbf{121}, 060201
(2018).

\bibitem{Krajenbrinketal2018} A. Krajenbrink, P. Le Doussal and S.
Prolhac, Nucl. Phys. B \textbf{936}, 239  (2018).

\bibitem{SMS2018} N. R. Smith, B. Meerson and P. V. Sasorov, J. Stat.
Mech. (2018) 023202.

\bibitem{GueudreLeDoussal2012} T. Gueudré and P. Le Doussal, Europhys.
Lett. \textbf{100}, 26006 (2012).

\bibitem{Borodin2016} A. Borodin, A. Bufetov, and I. Corwin, Annals
of Phys. \textbf{368}, 191 (2016).

\bibitem{Barraquand2017} G. Barraquand, A. Borodin, I. Corwin, and
M. Wheeler, arXiv:1704.04309.

\bibitem{CorwinShen2018} I. Corwin and H. Shen, Comm. Pure Appl. Math. \textbf{71},  2065 (2018).

\bibitem{ItoTakeuchi2018} Y. Ito and K. A. Takeuchi, Phys. Rev. E
\textbf{97}, 040103(R) (2018).

\bibitem{Krajenbrink2018} A. Krajenbrink and P. Le Doussal, SciPost Phys. \textbf{5}, 032 (2018).

\bibitem{MV2018} B. Meerson and A. Vilenkin, Phys.  Rev.  E
\textbf{98}, 032145 (2018).

\bibitem{Mikhailov1991} A. S. Mikhailov, J. Phys. A \textbf{24},
L757 (1991).

\bibitem{GurarieMigdal1996} V. Gurarie and A. Migdal, Phys. Rev.
E \textbf{54}, 4908 (1996).

\bibitem{Fogedby1998} H.C. Fogedby, Phys. Rev. E \textbf{57}, 4943
(1998).

\bibitem{Fogedby1999} H.C. Fogedby, Phys. Rev. E \textbf{59}, 5065
(1999).

\bibitem{Nakao2003} H. Nakao and A. S. Mikhailov, Chaos \textbf{13},
953 (2003).

\bibitem{Fogedby2009} H.C. Fogedby and W. Ren, Phys. Rev. E \textbf{80},
041116 (2009).

\bibitem{KMSparabola} A. Kamenev, B. Meerson, and P. V. Sasorov,
Phys. Rev. E \textbf{94}, 032108 (2016).

\bibitem{MSV_3d} B. Meerson, P. V. Sasorov and A. Vilenkin, J. Stat.
Mech. (2018) 053201.

\bibitem{EdwardsWilkinson}S. F. Edwards and D. R. Wilkinson, Proc.
R. Soc. Lond. A \textbf{381}, 17 (1982).

\bibitem{ChernykhStepanov} A. I. Chernykh and M. G. Stepanov, Phys.
Rev. E 64, 026306 (2001).


\bibitem{sameresult} The same scaling behavior can be obtained by returning to the dimensional
variables and demanding that the large deviation function
\eqref{eq: physical variables P(H,L,t)} be independent of the diffusion
coefficient $\nu$  \cite{SMS2018}.

\bibitem{ZR} Ya. B. Zel'dovich and Yu. P. Raizer, \textit{Physics of Shock Waves and High-Temperature Hydrodynamic Phenomena} (Academic Press, New York, 1966), vol. 1, p. 4.

\bibitem{Wolfram} Wolfram Research, Inc.,\\
\begin{footnotesize} http://functions.wolfram.com/ElementaryFunctions/ArcTan2/
\end{footnotesize}
\bibitem{Mazumder} S. Mazumder, \textit{Numerical Methods for Partial
Differential Equations} (Academic Press, New York, 2016).

\bibitem{LandauLifshitzFluidMechanics} L. D. Landau and E. M. Lifshitz,
\textit{Fluid Mechanics} (Reed, Oxford, 2000).

\bibitem{verified} We also verified our numerical results by checking the relation between $H$, $\ell$ and $f_+$, which follows from Eq. \eqref{eq: Naftali's relation} and the scaling relations \eqref{eq: hydrodynamic rescaling} and \eqref{eq: action scaling H->infty}.

\bibitem{Canet2011} L. Canet, H. Chat\'{e}, B. Delamotte, and N. Wschebor, Phys. Rev.
E \textbf{84}, 061128 (2011); \textbf{86}, 019904(E) (2012).

\end{thebibliography}
\end{document}